\documentclass[fleqn,usenatbib,useAMS]{mnras}
\usepackage[dvipsnames]{xcolor}
\usepackage{tikz,hyperref}
\usepackage{xcolor}
\usepackage{orcidlink}

\newcommand{\hi}{H\,\textsc{i}}
\newcommand{\hei}{He\,\textsc{i}}
\newcommand{\hii}{H\,\textsc{ii}}
\newcommand{\heii}{He\,\textsc{ii}}

\newcommand{\oi}{[O\,\textsc{i}]}
\newcommand{\ini}{[N\,\textsc{i}]}
\newcommand{\oii}{[O\,\textsc{ii}]}
\newcommand{\oiii}{[O\,\textsc{iii}]}
\newcommand{\oiv}{[O\,\textsc{iv}]}
\newcommand{\oiirls}{O\,\textsc{ii}}
\newcommand{\nii}{[N\,\textsc{ii}]}
\newcommand{\neiii}{[Ne\,\textsc{iii}]}

\newcommand{\nev}{[Ne\,\textsc{v}]}

\newcommand{\cliii}{[Cl\,\textsc{iii}]}

\newcommand{\sii}{[S\,\textsc{ii}]}
\newcommand{\siii}{[S\,\textsc{iii}]}
\newcommand{\ariii}{[Ar\,\textsc{iii}]}
\newcommand{\ariv}{[Ar\,\textsc{iv}]}

\newcommand{\ciii}{C\,\textsc{iii}}

\definecolor{lime}{HTML}{A6CE39}
\DeclareRobustCommand{\orcidicon}{
	\begin{tikzpicture}
	\draw[lime, fill=lime] (0,0) 
	circle [radius=0.13] 
	node[white] {{\fontfamily{qag}\selectfont \tiny ID}};
	\draw[white, fill=white] (-0.0625,0.095) 
	circle [radius=0.007];
	\end{tikzpicture}
	\hspace{-2mm}
}

\foreach \x in {A, ..., Z}{\expandafter\xdef\csname orcid\x\endcsname{\noexpand\href{https://orcid.org/\csname orcidauthor\x\endcsname}
			{\noexpand\orcidicon}}
}
\usepackage{newtxtext,newtxmath}

\usepackage{booktabs}
\usepackage[T1]{fontenc}
\usepackage{ae,aecompl}

\usepackage{graphicx}	% Including figure files
\usepackage{amsmath}	% Advanced maths commands
\usepackage{booktabs}
\usepackage{times}
\input{epsf}
\title[SDSS-V LVM: The Helix Nebula]{SDSS-V LVM: Revealing the Physical and Chemical Structure of the Helix Nebula}

\author[R. Orozco-Duarte et al.]{R. Orozco-Duarte$^{1}$\thanks{E-mail: r.orozco@irya.unam.mx}\orcidlink{0000-0003-2193-3005},
J. E. Méndez-Delgado$^{2}$\thanks{E-mail: jmendez@astro.unam.mx}\orcidlink{0000-0002-6972-6411},
J. A. Toalá$^{1}$\orcidlink{0000-0002-5406-0813},
L. Sabin$^{3}$\orcidlink{0000-0003-0242-0044},
C. Morisset$^{5}$\orcidlink{0000-0001-5801-6724},
S. F. Sánchez$^{2,4}$\orcidlink{0000-0001-6444-9307},
\and
H.~Ibarra-Medel$^{2}$\orcidlink{0000-0002-9790-6313},
W. J. Henney$^{1}$,\orcidlink{0000-0001-6208-9109},
A. Z. Lugo-Aranda$^{3,6}$\orcidlink{0000-0001-9226-9178},
A. Singh$^{8}$ \orcidlink{0009-0000-3962-103},
E. J. Johnston$^{9}$\orcidlink{0000-0002-2368-6469},
A. Roman-Lopes$^{11}$\orcidlink{0000-0002-1379-4204},
\and
C. G. Román-Zúñiga$^{3}$,\orcidlink{0000-0001-8600-4798},
I. Cruz-Gonzalez$^{2}$\orcidlink{0000-0002-2653-1120},
Guy S. Stringfellow$^{17}$\orcidlink{0000-0003-1479-3059},
M. Pe\~na$^{2}$\orcidlink{0009-0007-5891-420X},
A. Wofford$^{3}$,\orcidlink{0000-0001-8289-3428},
\and
E. Egorova$^{12}$\orcidlink{0000-0003-2717-8784},
O. Aranguré$^{3}$\orcidlink{0009-0004-7070-1782},
L. C. Castañeda-Carlos$^{2}$\orcidlink{0009-0000-2341-9865},
S. Torres-Peimbert$^{2}$\orcidlink{0000-0001-6775-8615},
L. Hernández-Martínez$^{2, 13}$\orcidlink{0000-0002-8274-5094}, 
\and
J. R. Brownstein$^{7}$\orcidlink{0000-0002-8725-1069},
R. de J. Zermeño$^{2}$\orcidlink{0009-0009-0081-4323},
I. Yu.~Katkov$^{14,15,16}$\orcidlink{0000-0002-6425-6879}, 
Oleg~V.~Egorov$^{12}$\orcidlink{0000-0002-4755-118X}, 
A. J. Mejía-Narváez$^{10}$\orcidlink{0000-0002-8931-2398},
\and
G. A. Blanc$^{18, 10}$\orcidlink{0000-0003-4218-3944},
Mónica A. Villa-Durango$^{3}$ \orcidlink{0009-0008-1605-4771},
\\\\
$^{1}$Instituto de Radioastronomía y Astrofísica, Universidad Nacional Autónoma de México, 58090 Morelia, Michoacán, México\\
$^{2}$Universidad Nacional Autónoma de México. Instituto de Astronomía. A.P. 70-264, 04510. Ciudad de México, México.\\
$^{3}$Universidad Nacional Autónoma de México. Instituto de Astronomía. A.P. 106, 22800. Ensenada, B.C. , México\\
$^{4}$Instituto de Astrofísica de Canarias, Vía Láctea s/n, 38205, La Laguna, Tenerife, Spain\\
$^{5}$Instituto de Ciencias Físicas, Universidad Nacional Autónoma de México, Av. Universidad s/n, 62210 Cuernavaca, Morelos, Mexico\\
$^{6}$Institute of Astrophysics, Facultad de Ciencias Exactas, Universidad Andrés Bello, Sede Concepción, Talcahuano, Chile\\
$^{7}$Department of Physics and Astronomy, University of Utah, 270 S. 1400 E. \#E2108, Salt Lake City, UT 84112, USA\\
$^{8}$Departamento de Astronom\'{i}a, Universidad de Chile, Camino del Observatorio 1515, Las Condes, Santiago, Chile\\
$^{9}$Instituto de Estudios Astrof\'isicos, Facultad de Ingenier\'ia y Ciencias, Universidad Diego Portales, Av. Ej\'ercito Libertador 441, Santiago, Chile.\\
$^{10}$Universidad de Chile, Av. Libertador Bernardo O'Higgins 1058, Santiago de Chile.\\
$^{11}$Department of Astronomy, Universidad de La Serena, Av. Raul Bitran 1302, La Serena, Chile.\\
$^{12}$Astronomisches Rechen-Institut, Zentrum f\"ur Astronomie der Universit\"at Heidelberg, M\"onchhofstr. 12-14, D-69120 Heidelberg, Germany.\\
$^{13}$Universidad Nacional Aut\'onoma de M\'exico, Facultad de Ciencias, Av. Universidad 3000,  04510, Cd. de M\'exico\\
$^{14}$New York University Abu Dhabi, PO Box 129188, Abu Dhabi, UAE\\
$^{15}$Center for Astrophysics and Space Science (CASS), New York University Abu Dhabi, PO Box 129188, Abu Dhabi, UAE\\
$^{16}$Sternberg Astronomical Institute, Lomonosov Moscow State University, Universitetskij pr., 13,  Moscow, 119234, Russia\\
$^{17}$University of Colorado, Boulder, CO 80309, USA\\
$^{18}$Observatories of the Carnegie Institution for Science, 813 Santa Barbara Street, Pasadena, CA 91101, USA\\
}
\date{\today}%Accepted XXX. Received YYY; in original form ZZZ}

\pubyear{2025}

\begin{document}
\label{firstpage}
\pagerange{\pageref{firstpage}--\pageref{lastpage}}
\maketitle

\begin{abstract}
We present a spatially contiguous spectroscopic analysis of the physical and chemical structure of the Helix Nebula (NGC~7293, PNG~036.1-57.1) using integral-field observations from the SDSS-V Local Volume Mapper (LVM). The nearly complete nebular coverage allows us to map extinction, electron density and temperature, ionisation structure, and chemical abundances within a common dataset. We reconstruct calibrated datacubes from the LVM row-stacked spectra and measure 41 optical emission lines from hydrogen, helium, and collisionally excited ions. The Helix is globally a low-density nebula, with characteristic electron densities of $\sim10^{2}$~cm$^{-3}$, but shows a strongly structured thermal distribution. $T_{\rm e}([\mathrm{O\,III}])$ exhibits a U-shaped radial profile, decreasing from the central cavity to intermediate radii and rising toward the outer nebula, while $T_{\rm e}([\mathrm{N\,II}])$ increases outward and $T_{\rm e}([\mathrm{O\,II}])$ decreases. These contrasting behaviours show that different diagnostics trace distinct ionisation zones. The spatially resolved abundance analysis yields a near-solar oxygen abundance, $12+\log(\mathrm{O/H})\simeq8.7$, consistent with an approximately chemically homogeneous nebula once ionisation structure is considered. Simulated long-slit extractions show that inferred O/H depends systematically on slit orientation and can remain biased after ionisation correction, demonstrating the importance of spatial coverage in strongly stratified nebulae. We also find sulfur systematically deficient relative to other $\alpha$-elements, consistent with the planetary-nebula sulfur anomaly. These results highlight the power of wide-field integral-field spectroscopy for linking the thermal, ionisation, and chemical structure of extended planetary nebulae.
\end{abstract}

% Select between one and six entries from the list of approved keywords.
% Don't make up new ones.
% Reset the two-column page geometry after the long title/author block
\makeatletter
\global\@colht=\textheight
\global\@colroom=\textheight
\global\vsize=\textheight
\makeatother
\clearpage
%%%%%%%%%%%%%%%%%%%%%%%%%%%%%%%%%%%%%%%%%%%%%%%%%%

%%%%%%%%%%%%%%%%% BODY OF PAPER %%%%%%%%%%%%%%%%%%

\section{Introduction}
\label{sec:intro}
Planetary nebulae (PNe) represent one of the final evolutionary stages of low- and intermediate-mass stars ($M_\mathrm{ZAMS} \lesssim 1-8$ M$_\odot$)\footnote{ZAMS stands for zero age main sequence.}. These objects arise when such stars shed their outer layers towards the end of their lives by evolving through the asymptotic giant branch (AGB) phase \citep{Habing1996, Herwig2005}. Subsequently, these stars become post-AGB objects, developing a strong ionising photon flux and a fast wind \citep{Vassiliadis1994, Blocker1995}. These two effects combine to compress and ionise the previously ejected material, giving birth to a PN \citep[see][and references therein]{Kwitter2022, Kwok2024}. However, the shaping mechanism and resulting morphologies of PNe remain topics of active debate, with binary interactions and magnetic fields often invoked as additional factors \citep{Nordhaus2006, DeMarco2009, Jones2017}.

\begin{figure*}
    \centering
    \includegraphics[width=0.7\linewidth]{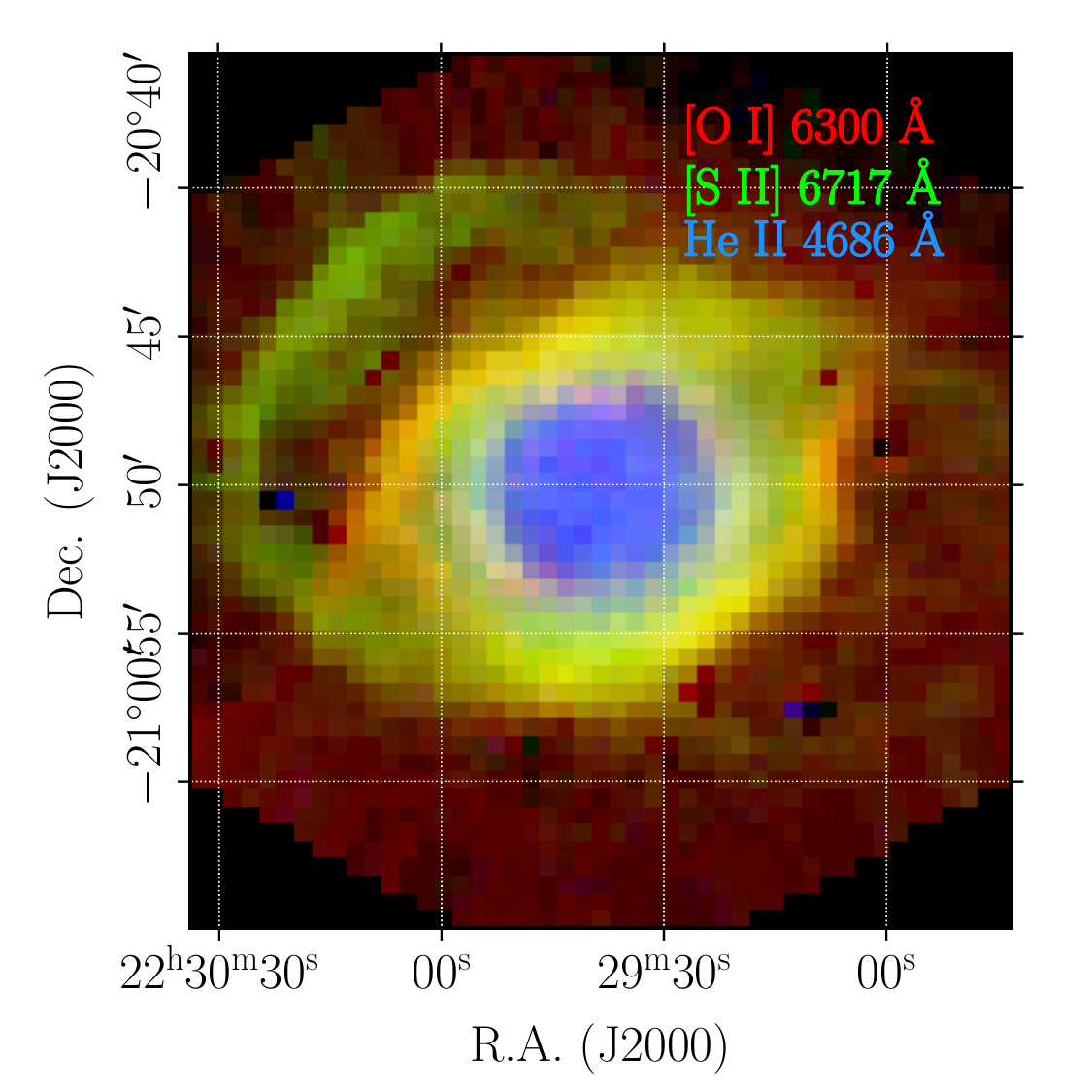}
    \caption{RGB composite of the Helix Nebula highlighting its ionisation structure. The red, green, and blue channels correspond to \oi~$\lambda6300$, \sii~$\lambda6716$, and \heii~$\lambda4686$, respectively. The image reveals a clear stratification from neutral and low-ionisation gas in the outer regions to highly ionised material concentrated in the central cavity.}
    \label{fig:helix_iostructure}
\end{figure*}

The Helix Nebula (a.k.a. NGC~7293, PNG 036.1-57.1) is a nearby \citep[$d\approx200$ pc;][]{Harris2007,BailerJones2021} and extended PN. Its brightest nebular emission has an angular size of $12\times20$~arcmin$^{2}$ \citep{ODell1998,ODell2004}. However, fainter emission extends up to $\sim32$~arcmin in the east--west direction, as revealed by deep optical, UV, and IR observations \citep{Meaburn2013,vandeSteene2015}. The central star of this PN is WD~2226$-$210, a hot DAO-type white dwarf with an effective temperature of 120,000\,K and surface gravity \(\log g = 7.2 \) \citep{Traulsen2005, Filiz2024}. Its effective temperature and evolved state place it well in the post-AGB phase where the stellar remnant emits strong UV photons. This radiation field plays a key role in shaping the observed morphology of the nebula and in sustaining the observed emission-line spectrum. 

Given its large angular size, {\it Hubble Space Telescope} is not able to obtain complete images of the Helix Nebula. However, it has been used to peer into the neutral and ionised gas properties of the globules projected towards the central regions of this PN \citep[e.g.,][]{ODell2005,Meixner2005}. Not only are the globules H${2}$-rich, the Helix Nebula has been known to harbour complex molecules such as HCN, HCO$^{+}$, and H$_{2}$CO \citep[e.g.,][]{Bachiller1997,Zack2013,Etxaluze2014}. These molecular components are thought to survive in dense clumps shielded from the central radiation field \citep{Huggins2002}.

Abundance determinations of the Helix Nebula have been performed by several authors, although most studies have focused on different regions within the brightest regions in the nebula through long slit spectroscopy \citep{Hawley1978,Zuckerman1986, Manchado1989,Peimbert1995,Henry1999}. For spatially structured nebulae, such measurements can be sensitive to the position and orientation of the spectroscopic aperture, as demonstrated by IFU studies using simulated pseudo-slits \citep{Akras2022}. In the era of integral field spectroscopy (IFS), different teams have obtained resolved maps of the ionisation structure, abundances, and kinematics within PNe \citep[e.g.,][]{GarciaRojas2016,MonrealIbero2020,Guerrero2021,RechyGarcia2021,RamosLarios2022,GomezLlanos2024,MontoroMolina2025}. These studies have demonstrated the importance of spatially resolved analyses in revealing chemical inhomogeneities and ionisation structure \citep{Tsamis2008}. Although the Helix Nebula is one of the most iconic ionised nebula in the sky, there is still no spatially resolved optical spectroscopic analysis of it, which would provide a comprehensive mapping of its physical characteristics and chemical abundances. 

The Sloan Digital Sky Survey V (SDSS-V) is a panoptic spectroscopic survey designed to map the sky with multi-object spectroscopy across both hemispheres, complemented by integral-field spectroscopy mainly in the southern hemisphere \citep{Kollmeier2026}. As part of SDSS-V, the Local Volume Mapper (LVM) is a new wide-field integral-field spectroscopy survey designed to map ionised gas and stellar populations across large angular scales in the Milky Way and nearby systems (\citealt{Drory2024}; Blanc et al., in prep.). The combination of contiguous spatial coverage and spectroscopic resolution opens a new observational window for studying extended nebulae with fully resolved physical diagnostics. This capability is particularly critical for nearby, spatially extended nebulae such as the Helix \citep{Kreckel2024,Kreckel2026,Villa-Durango2025,Sarbadhicary2026,Sanchez2026,Singh:2026,Hilder2026,Sattler:2026}. 

Recent SDSS-V Data Release 19 \citep{SDSS2025COL} preview observations include an integral-field unit pointing of the Helix Nebula \citep[see][]{Sanchez2026}, providing spatially continuous spectroscopy over a substantial region of this nearby PN. \citet{Sanchez2026} presented the corresponding LVM-DAP data products and demonstrated that the LVM observations recover the main structural properties previously inferred from classical aperture and imaging studies, including the ionisation stratification from the compact He$^{2+}$ core to the bright \oiii\ ring and the extended low-ionisation envelope. Their analysis also highlighted the low dust content, the overall homogeneity of the physical conditions across the main ring, and ionised-gas kinematics consistent with a slowly expanding, limb-brightened shell. 

The present work extends this initial analysis by combining the three available exposures and reconstructing a dedicated datacube for a detailed nebular analysis. Rather than relying on the standard LVM-DAP products, we independently measure the emission-line spectrum in each reconstructed spaxel and use these measurements to derive spatially resolved extinction, multiple electron-temperature and density diagnostics, ionisation structure, ionic and total chemical abundances, and ionisation correction factors. This enables us to investigate the radial and two-dimensional variations of the physical and chemical conditions across the nebula, including temperature structures and abundance variations that were not addressed in the initial LVM analysis. The spatially complete coverage also allows us to evaluate explicitly the effects of spatial sampling on abundance determinations through simulated long-slit observations.

This paper is organised as follows. Section~\ref{sec:obs} describes the observations, data reduction, and datacube reconstruction. Section~\ref{sec:spectral_analysis} presents the spectral analysis and the determination of the physical and chemical conditions of the gas. Section~\ref{Sec:ionisation_structure} discusses the global ionisation structure of the nebula, while Section~\ref{Sec:reddening} presents the reddening map. Sections~\ref{sec:density_determinations} and~\ref{sec:temperature_determinations} describe the determination of the electron density and temperature, respectively. Section~\ref{sec:chemical_abundances} presents the chemical abundance analysis, and Section~\ref{sec:integrated_spectrum} provides an integrated analysis of the nebula. Finally, Section~\ref{sec:discussion} discusses our results, and Section~\ref{sec:conclusions} summarises the main conclusions.

\section{Observations}
\label{sec:obs}

% Figure~\ref{fig:helix_iostructure} illustrates the footprint of the observations.

The LVM instrument consists of four 16\,cm telescopes installed at Las Campanas Observatory, Chile \citep{Herbst2024}. Each telescope is equipped with a fibre-fed integral-field unit: one dedicated to science, two to sky subtraction, and one to the acquisition of spectrophotometric standards \citep{Drory2024}. The fibres feed three DESI-type spectrographs, each with three channels (blue, red, and infrared), covering the $3600$--$9800$\,\AA\ range at a spectral resolution of $R\simeq4000$ at H$\alpha$, corresponding to a velocity resolution of $\sim75$\,km s$^{-1}$. The science IFU comprises 1801 fibres arranged in a hexagonal pattern, providing a contiguous field of view of $\sim30$\,arcmin, which is particularly well suited for mapping nearby extended nebulae \citep{Drory2024, Kreckel2024, Villa-Durango2025, Sarbadhicary2026, Singh:2026, Sattler:2026}.

We observed the Helix Nebula in September 2023, obtaining three 900\,s exposures at the same position (no dithering) with the LVM science IFU under photometric conditions, ensuring stable atmospheric transmission during the observations. Because of its high surface brightness and large apparent size, the $\sim$30\,arcmin LVM field of view fully encloses the brightest regions of the nebula up to a minimum H$\alpha$ flux of $2.22 \times 10^{-17}$ erg\,s$^{-1}$\,cm$^{-2}$ for an S/N $\approx$ 2.5, although this sensitivity limit varies across the field due to instrumental and background variations. Figure~\ref{fig:helix_iostructure} illustrates the footprint of the Helix Nebula observations derived from the LVM data, highlighting the wide-field, spatially contiguous coverage achieved across the nebula.

\subsection{Data reduction}
The raw LVM exposures are processed with the standard LVM data-reduction pipeline in its version 1.2.1 (A.~Mejía-Narváez, in preparation), following procedures broadly consistent with modern IFU reduction frameworks \citep{Law2016}. In brief, following \cite{Sanchez2026}, the pipeline applies basic detector cleaning and calibration, extracts the fibre spectra, performs wavelength and flux calibration, subtracts the sky contribution using the dedicated sky fibres, and attaches an astrometric solution. The wavelength calibration is based on arc-lamp exposures, while flux calibration relies on observations of standard stars obtained simultaneously with science exposures. The final product of the LVM-DRP is a calibrated row-stacked spectra (RSS) FITS file, in which each row corresponds to a science fibre, together with propagated error estimates, quality masks, and the associated astrometric information. 
\textbf{The statistical uncertainties are propagated pixel by pixel throughout the LVM-DRP, including the sky-subtraction and flux-calibration stages, assuming Poisson statistics and accounting for the detector noise. These uncertainties are subsequently propagated during the cube reconstruction and provide the formal wavelength-dependent errors used in our emission-line analysis. Systematic uncertainties associated with the absolute flux calibration are not treated separately in our analysis. The propagation does not explicitly account for covariance between neighbouring spectral samples introduced during the resampling, and the reported uncertainties should therefore be interpreted as formal per-pixel errors.}
These reduced RSS spectra constitute the starting point for the cube reconstruction and subsequent emission-line analysis performed in this work. For further details on the reduction procedure, we refer the reader to \citet{Sanchez2025}.

\subsection{Cube reconstruction}
We reconstructed three-dimensional datacubes from the LVM RSS using the \texttt{3DCubeGen}\footnote{\url{https://github.com/hjibarram/3DCubeGen}} package \citep{Ibarra-Medel2026}. The code performs a 2D spatial interpolation of the fibre fluxes onto a regular WCS grid by distributing the flux of each fibre with a Gaussian kernel centred on the fibre position. The kernel is defined by distributing the nominal fibre diameter, but its width is a tunable reconstruction parameter rather than being fixed to the full fibre diameter. The output pixel scale is also defined in terms of the fibre diameter in order to control the spatial sampling of the reconstructed cube.

For each wavelength channel, the fibre contributions overlapping a given spaxel are combined using inverse-variance weighting, and known bad or low-quality fibres are masked using the provided slitmaps. Alongside the flux cube, the pipeline propagates uncertainties at every step to deliver a co-registered error cube. The resulting FITS product adopts the standard $(\lambda, y, x)$ axis convention and include WCS metadata.

Such reconstruction methods inherently introduce spatial variance between adjacent spaxels, which should be taken into account in subsequent analyses \citep{Husemann2013}.
To mitigate the excessive propagation of stellar absorption features from bright field stars, while also avoiding the introduction of artificial structures due to oversampling, the cube reconstruction algorithm of \cite{Ibarra-Medel2026} adopts a set of parameters motivated by the general considerations for fibre-fed IFU cube reconstruction discussed by \citet{Sanchez2012}. For the present cube, these correspond to a Gaussian kernel width set to one third of the fibre diameter, a spatial pixel scale equal to the fibre diameter, and a kernel shape factor of $2\arcsec$. This configuration represents a compromise between spatial resolution and noise correlation, optimised for extended low-surface-brightness emission.

Although the three exposures were obtained at the same pointing without spatial dithering, the adopted reconstruction was deliberately chosen to minimise additional spatial interpolation: the output spaxel size is one fibre diameter, while the Gaussian kernel width is only one third of this scale. For the LVM fibre diameter, this corresponds to a projected scale of about 0.034 pc at the adopted distance of 200 pc. A detailed sensitivity analysis of the 3DCubeGen reconstruction parameters is presented by \cite{Ibarra-Medel2026}, who show that decreasing the interpolation kernel below the fibre scale does not recover correspondingly finer independent spatial information, since the effective resolution eventually becomes limited by the fibre aperture, whereas larger kernels mainly introduce additional spatial smoothing. In the particular case of the Helix, the physical impact of this trade-off is limited by its proximity, because reconstruction effects operating on approximately one fibre scale correspond to only a few hundredths of a parsec. The reconstruction therefore does not strongly over-sample the native fibre information. Moreover, the physical conditions and ionisation maps discussed in this work exhibit coherent structures extending over several fibre scales, rather than being defined by individual spaxel-to-spaxel fluctuations. We therefore regard the latter primarily as a limitation on the interpretation of the smallest spatial scales, while the large-scale radial and morphological trends on which our conclusions are based remain robust.
% To mitigate the excessive propagation of stellar absorption features from bright field stars, while also avoiding the introduction of artificial structures due to oversampling, the procedure follows a similar approach to one presented in \citep{Sanchez2012}. We adopt a specific set of reconstruction parameters: a Gaussian kernel width set to one third of the fiber diameter, a spatial pixel scale equal to the fiber diameter, and a kernel shape factor of $2\arcsec$.

\subsection{The integrated optical spectrum of the Helix Nebula}
\label{sec:integrated_spectrum_data}
Following the datacube reconstruction, we constructed an integrated optical spectrum of the Helix Nebula within the LVM footprint. This spectrum provides a global characterisation of the nebular emission and serves as a reference for identifying the main emission lines used in the subsequent spatially resolved analysis.

\begin{figure*}
\centering
\includegraphics[width=1.0\linewidth]{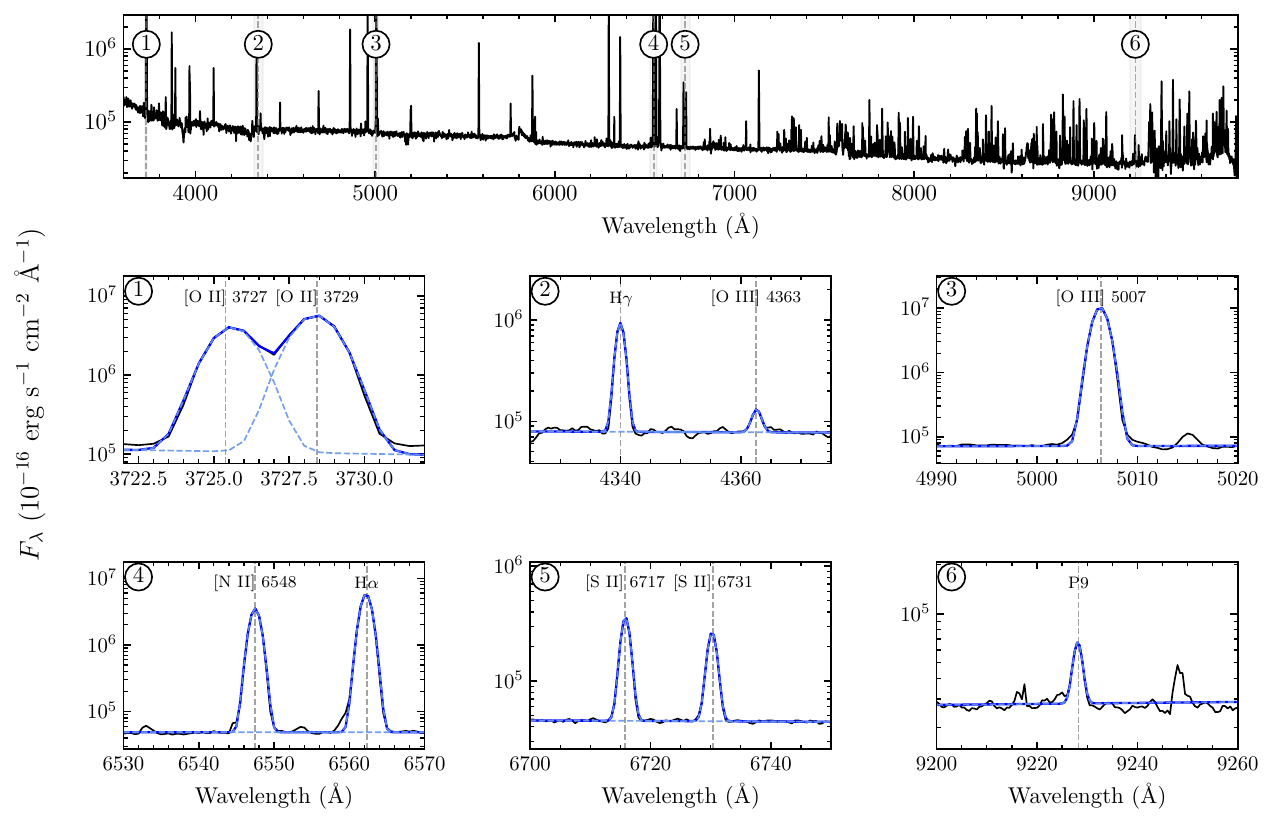}
\caption{Integrated optical spectrum of the Helix Nebula obtained by summing all valid spaxels within the LVM footprint. The top panel shows the full optical wavelength coverage of the LVM data, while the numbered lower panels present zoom-in regions around selected diagnostic features. The black curves show the integrated spectrum and the blue curves show Gaussian fits to the emission lines. Prominent lines are labelled in each zoom panel, including the \oii~doublet, the auroral \oiii~$\lambda4363$ line, strong \oiii~and H$\alpha$ emission, the \sii~doublet, and the P9 recombination line. Residual sky features are present mainly at red wavelengths, but lines affected by sky-subtraction artefacts were excluded from the quantitative analysis. Wavelength is given in \AA.}
\label{fig:total_spectrum}
\end{figure*}

Figure~\ref{fig:total_spectrum} presents the integrated spectrum obtained by summing the flux from all valid fibres across the data. The resulting spectrum displays a rich variety of emission features over the full LVM wavelength coverage, including strong hydrogen and helium recombination lines, together with numerous collisionally excited lines from low-, intermediate-, and high-ionisation species. The simultaneous presence of these transitions reflects the broad ionisation structure of the Helix Nebula and provides an excellent dataset for both spatially resolved and integrated nebular analyses, while retaining a direct link to the spatially resolved measurements.

An important advantage of the integrated spectrum is that it allows the application of standard ICF schemes, such as those proposed by \citet{Delgado-Inglada:2014}. These prescriptions are based on photoionisation models that implicitly assume the inclusion of the full ionisation structure of the nebula. Spatially integrated observations are therefore more directly comparable to the model framework on which such relations are calibrated than individual spaxels sampling only restricted ionisation zones. Nevertheless, integrated spectra also compress intrinsically stratified gas into a single luminosity-weighted measurement, which must be considered when interpreting the derived physical conditions.

It is worth noting that some residual sky features remain visible in portions of the spectrum, particularly at redder wavelengths, owing to the difficulty of achieving a perfect subtraction of strong atmospheric emission lines. Nevertheless, special care was taken during the line-measurement stage to avoid affected features. Lines showing clear contamination, uncertain continua, or severe blending with residual sky emission were excluded from the quantitative analysis, ensuring that the derived physical conditions and abundances are based only on reliable measurements.

\section{Spectral analysis}
\label{sec:spectral_analysis}
All emission-line measurements were performed on the reconstructed datacube, ensuring full spatial consistency across the field of view, using our own line-fitting procedure rather than the LVM DAP products \citep{Sanchez2025}. We first inspected the global spectrum to identify the brightest features, including the main hydrogen recombination lines, \hei\, and \heii\, lines, and strong collisionally excited lines such as \oii, \oiii, \sii, \siii, which were then used to define the full list of the lines to be fitted. Owing to the broad spectral coverage of LVM, we were able to observe simultaneously a large number of strong forbidden and permitted emission lines, spanning a wide range of ionisation potentials.

Although stellar absorption features in some \hi~lines may be present in limited regions of the nebula \citep{Sanchez2026}, they are not dominant given the nature of the object, which is primarily governed by nebular emission ionised by the central stellar remnant. Nevertheless, a non-negligible stellar continuum contribution is detected in a subset of spaxels across the field of view, particularly in regions affected by foreground and background stars.

Under these conditions, we adopt a line-by-line approach to measure the emission features of interest, Table~\ref{tab:emission_lines} summarises the full list of 41 emission lines that were analysed by this approach. For each transition, the local continuum is estimated on both sides of the line and fitted with a linear function, which is subsequently subtracted from the spectrum. The residual emission is then fitted with a Gaussian profile within a predefined wavelength interval, following standard spectroscopic practices \citep{Tresse1999}. The Gaussian amplitude, centroid, and width are allowed to vary during the fit, and the integrated line flux is obtained from the best-fitting profile. The local noise level is estimated from the dispersion of the adjacent continuum regions and is used both in the fitting procedure and to estimate the line signal-to-noise ratio. This same local noise estimate is adopted as the uncertainty in the Gaussian line fit and is propagated to the integrated line-flux uncertainty. In this way, fluctuations in the local continuum are included in the formal line-flux errors, although the covariance of the fitted continuum itself is not explicitly propagated. This strategy allows for a careful and robust determination of line fluxes, particularly for weak auroral transitions such as \oiii~$\lambda4363$, \nii~$\lambda5755$, and \siii~$\lambda6312$. These lines are critical for the derivation of electron temperature and are especially sensitive to systematic uncertainties arising from even small inaccuracies in the continuum subtraction \citep{Osterbrock:2006}. For each spaxel, all detected emission lines are fitted with single Gaussian profiles to derive their integrated fluxes and kinematic properties. To improve convergence, the fitting procedure is repeated using a small range of initial Gaussian widths when necessary, and fits that do not satisfy basic flux, line-width, and signal-to-noise quality criteria are rejected. In the case of the \oii~$\lambda\lambda 3726,3729$ lines, there is a partial blend between them due to the spectral resolution of the LVM. These lines are therefore fitted simultaneously with two Gaussian components, using separate wavelength intervals around each transition to initialize their centroids and amplitudes. The lower panels presented in Fig.~\ref{fig:total_spectrum} reveal some hydrogen recombination lines from the Balmer series, well-known forbidden transitions such as  \oiii~$\lambda5007$ and the \sii~$\lambda\lambda6716,6731$ lines, and many others, confirming the multi-phase ionisation structure of the nebula.

\subsection{Determining the nebular physical and chemical conditions}
\label{sec:determining_physical_conditions}

To derive the physical conditions and ionic abundances, we used the {\texttt{PyNeb}} package \citep{Luridiana2015, Morisset:2020} in version 1.1.24, which is specifically designed for computing nebular diagnostics from optical emission lines. This tool provides a self-consistent framework for deriving plasma properties based on atomic data and emissivity calculations. Because extinction must be corrected before using the observed fluxes, we first derived a reddening map. For this, we used the \cite{Fitzpatrick1999} extinction law adopting an $R_V$ value of 3.1 to obtain the extinction coefficient $c(\mathrm{H}\beta)$ from hydrogen recombination line ratios. This choice is appropriate for diffuse interstellar conditions and is commonly adopted in Galactic nebular studies. Rather than relying on a single diagnostic, we computed $c(\mathrm{H}\beta)$ from multiple recombination-line pairs and inspected their consistency, reducing systematic uncertainties associated with individual ratios.

\begin{figure*}
    \centering
    \includegraphics[width=1.0\linewidth]{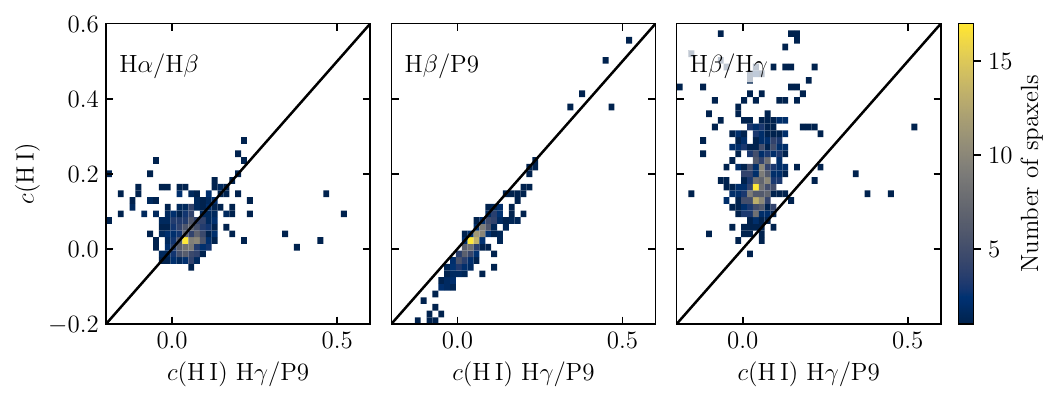}
    \caption{Comparison of the extinction coefficient $c(\mathrm{H}\beta)$ derived from multiple \hi~recombination-line ratios. Each panel shows the two-dimensional distribution for a different diagnostic pair, plotted against the reference value obtained from the H$\gamma$/P9 ratio. The colour scale indicates the number of spaxels in each bin, and the solid black line marks the one-to-one relation. The observed scatter reflects measurement uncertainties and possible systematic differences among the diagnostics, highlighting the importance of comparing multiple \hi~ratios to obtain a robust extinction estimate.}
    \label{fig:cHbeta_comparison}
\end{figure*}

Figure~\ref{fig:cHbeta_comparison} compares the extinction values derived from different line ratios against those obtained from H$\gamma$/P9. Although the various estimators broadly follow the same trend, significant scatter is present at the spaxel level, and some individual measurements yield formally negative values. For the final extinction map, we adopt the median of five independent estimates of $c(\mathrm{H}\beta)$ derived from H$\beta$/P9, H$\delta$/P9, H$\beta$/H$\delta$, H$\gamma$/P9, and H$\beta$/H$\alpha$. The use of several diagnostics reduces the sensitivity of the adopted extinction coefficient to any individual noisy or discrepant ratio. Residual wavelength-dependent uncertainties in the relative flux calibration may nevertheless contribute to the observed scatter, particularly for the Balmer-to-Paschen ratios, which span a large wavelength range. Only 111 out of 1153 valid spaxels (9.6 per cent) yield a formally negative median extinction coefficient. These values are consistent with statistical and calibration uncertainties around the very low extinction characteristic of the Helix rather than with physical negative reddening. See Section~\ref{Sec:reddening} for further discussion.

Once the de-reddened fluxes were obtained, we derived electron density and temperature maps using classical diagnostics based on collisionally excited lines (CELs). The electron density ($n_{\rm e}$) can be determined from line ratios involving transitions arising from similar atomic energy levels but with different collisional de-excitation rates, such as \cliii~$\lambda5517/\lambda5538$, \oii~$\lambda3726/\lambda3729$, and \sii~$\lambda6716/\lambda6731$. These diagnostics were measured across the field of view and used to infer the density structure of the Helix Nebula, probing different ionisation zones.

The density calculations were performed using the \texttt{getTemDen} routine from \textsc{PyNeb}, assuming an initial electron temperature of $9,000 \pm 2,000$ K within its Machine Learning module implemented in \textsc{ai4neb}\footnote{https://github.com/Morisset/AI4neb}. This temperature range accounts for potential spatial variations across the nebula. Although such variations have a negligible impact on the aforementioned density diagnostics due to the similar excitation energies of the involved transitions this assumption is validated in subsequent steps of the analysis.

The use of machine-learning-based regression routines significantly accelerates the computations, enabling the determination of electron density across all spaxels. In addition, uncertainties are propagated on a spaxel-by-spaxel basis through a Monte Carlo approach with 1000 realisations for each spaxel, while maintaining consistency with traditional diagnostic methods.

In a similar manner, the electron temperature ($T_{\rm e}$) was estimated using CEL ratios involving transitions arising from significantly different excitation energies, such as \nii~$\lambda5755/(\lambda6548+\lambda6583)$, \oiii~$\lambda4363/(\lambda4959+\lambda5007)$, and \siii~$\lambda6312/(\lambda9069+\lambda9531)$. In addition, temperature maps were derived from the \oii~$(\lambda7319+\lambda7320)/(\lambda3726+\lambda3729)$ and \sii~$\lambda4069/(\lambda6716+\lambda6731)$ ratios, allowing cross-validation between independent diagnostics. We note that the \sii~$\lambda4069$ feature may include a contribution from nearby permitted \ciii~lines at $\lambda\lambda4067.94,4068.91,4070.26$. This contamination is expected to have little impact on individual spaxel measurements, but may become more relevant in the integrated spectrum, potentially leading to a slight overestimate of the temperature derived from this diagnostic.

The auroral lines \oii~$\lambda7330+\lambda7331$ and \sii~$\lambda4076$, although detected in some cases, were not used for the temperature determination. The former is affected by contamination from nearby sky emission lines, which can artificially enhance the measured flux and lead to overestimated temperatures. In the latter case, the line is dominated by noise, as the ratio \sii~$\lambda4076/\lambda4069 \approx 0.33$ remains nearly constant over a wide range of densities below $10^{5}\ \mathrm{cm}^{-3}$, limiting its diagnostic power. Each of these diagnostics probes the thermal conditions of different ionisation zones, corresponding to low-, high-, and intermediate-ionisation regions \citep{OrteGarcia:2025}, and will be discussed in more detail in the following sections.

These temperatures were derived using the \texttt{getTemDen} routine from \textsc{PyNeb}, employing its Machine Learning implementation in \textsc{ai4neb}, following the same approach adopted for the determination of $n_{\rm e}$. As input for the temperature calculations, we used the electron density map derived from the \sii~$\lambda6716/\lambda6731$ ratio, which provides the most complete spatial coverage across the field of view. However, the impact of density on the derived temperature structure is negligible, as the nebula lies within the low-density regime ($n_{\rm e} < 10^3\ \mathrm{cm}^{-3}$), where these diagnostics are largely insensitive to density variations \citep{mendezDelgado2023b}, ensuring the robustness of the temperature estimates.

Finally, we derived ionic abundances using {\texttt{PyNeb}} by propagating the uncertainties through Monte Carlo simulations on the line fluxes, electron temperatures, and densities. We measured ionic abundances for He$^{+}$, He$^{2+}$, O$^{+}$, O$^{2+}$, N$^{+}$, S$^{+}$, S$^{2+}$, Ne$^{2+}$, Ar$^{2+}$, and Cl$^{2+}$. For each ion, multiple realisations were generated and the median and standard deviation of the resulting distributions were adopted as the ionic abundance and its associated uncertainty, minimising the impact of non-linear error propagation. Total elemental abundances were then computed by summing the observed ionic species of each element, with implicit assumptions on ionisation correction factors to be discussed in subsequent sections.

\section{Global ionisation structure}
\label{Sec:ionisation_structure}

To guide the interpretation of the spatially resolved properties of the Helix Nebula, we present in Figure~\ref{fig:helix_iostructure} an RGB composite highlighting the ionisation stratification across the nebula. The image combines emission from \oi~$\lambda6300$ (red), \sii~$\lambda6716$ (green), and \heii~$\lambda4686$ (blue), tracing neutral, low-ionisation, and highly ionised gas, respectively, thus providing a spatially resolved view of the ionisation structure \citep{Osterbrock:2006}. The central cavity is dominated by \heii~emission, in agreement with \cite{Sanchez2026} and indicates the presence of highly ionised gas exposed to a hard radiation field with energies higher than 54.6 eV, sufficient to produce He$^{2+}$. Such conditions imply the presence of high-ionisation species (e.g. O$^{3+}$), which are not directly traced by optical diagnostics \citep{Delgado-Inglada:2014}. Surrounding this region, a bright ring of \sii~emission traces denser, partially ionised material, while the outermost regions are characterised by enhanced \oi~emission, associated with neutral or transition zones.

This layered structure is consistent with expectations from photoionisation equilibrium, in which the ionisation parameter decreases with distance from the central star \citep{Stasinska2007}. The transition from the \heii-dominated emission to \sii~and \oi~regimes reflects the progression from fully ionised gas to partially ionised zones, where the radiation field is attenuated and neutral species persist.
This stratified structure reflects the radial variation of the ionising radiation field and provides a direct visual framework for interpreting the physical conditions and chemical abundances discussed in the following sections. In particular, the confinement of \heii~emission to the inner regions anticipates the presence of highly ionised species such as O$^{3+}$, not directly traced by optical diagnostics and would instead require ultraviolet or infrared lines, such as \oiv~25.9~$\mu$m.

\section{Reddening map}
\label{Sec:reddening}

\begin{figure}
    \centering
    \includegraphics[width=1.0\linewidth]{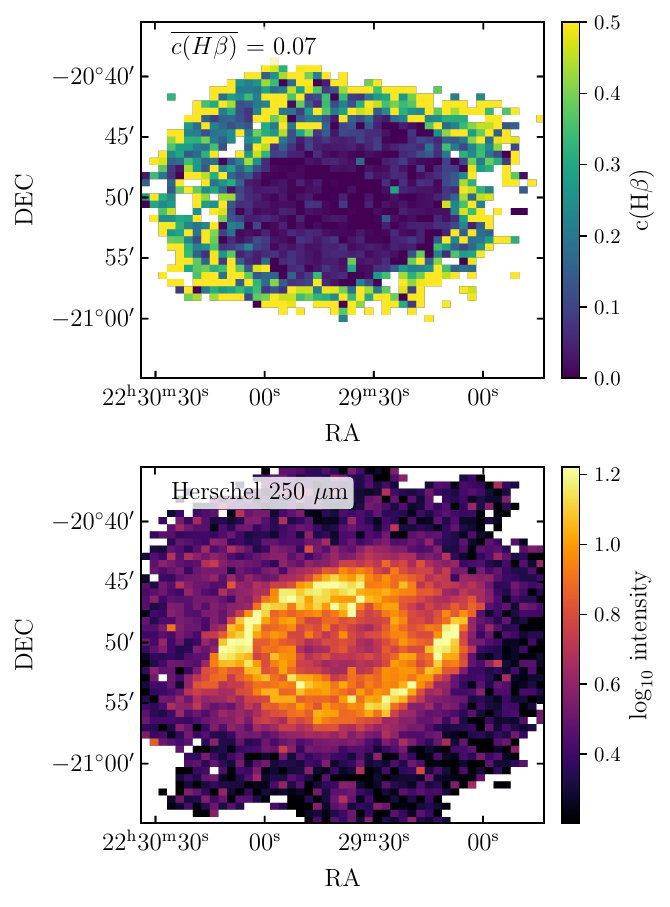}
\caption{(Top) Spatial distribution of the extinction coefficient $c(\mathrm{H}\beta)$ across the Helix Nebula derived from \hi~recombination lines. The map reveals very low extinction in the central cavity and enhanced values along the surrounding bright rim, outlining an oval-shaped shell consistent with the projected toroidal structure of the nebula. (Bottom) Herschel/SPIRE 250~$\mu$m emission, shown in logarithmic scale and reprojected onto the same spatial grid, for comparison with the extinction map.}  
\label{fig:extinction_map}
\end{figure}

Figure~\ref{fig:extinction_map} shows the spatial distribution of the extinction coefficient $c(\mathrm{H}\beta)$ derived from recombination lines (RLs) of \hi, providing a direct tracer of the dust attenuation affecting the ionised gas. The nebula exhibits very low extinction in the central regions, with typical values $c(\mathrm{H}\beta)\approx 0.0$--$0.1$. Moving outwards, the extinction slightly increases, reaching $c(\mathrm{H}\beta)\approx 0.2$--$0.3$ along the bright rim that encloses the central cavity. This radial trend indicates that dust is preferentially distributed in the outer regions of the ionised shell. The resulting morphology is an oval-shaped shell, consistent with the projected geometry of NGC\,7293 on the plane of the sky. The low extinction levels in the inner nebula are in agreement with \cite{ODell1998}, who reported little to negligible internal extinction in the Helix. The modest increase in $c(\mathrm{H}\beta)$ occurs mainly outside the bright optical rim, broadly coinciding with regions where \citet{vandeSteene2015} found enhanced H$_2$/H$\beta$ ratios, and is therefore consistent with attenuation associated with molecular and dusty material surrounding the ionised shell.

When comparing the extinction morphology with the dust continuum maps presented by \cite{vandeSteene2015}, we find that the far-infrared dust emission traces the clumpy ring-like structure surrounding the central cavity, as previously reported. However, the highest 250\,$\mu$m emission does not coincide with the highest values of $c(\mathrm{H}\beta)$ in our map. Instead, the spaxels associated with the strongest 250\,$\mu$m emission typically show modest extinction values, $c(\mathrm{H}\beta) \lesssim 0.2$. This suggests that the far-infrared emission is more closely associated with the dust-rich outer ring than with localised peaks in the optical extinction.

Nevertheless, the spatial overlap between the 250\,$\mu$m emission and regions of non-negligible extinction indicates that the dust traced in the far-infrared may still contribute to the line-of-sight attenuation of the ionised gas, rather than being confined to a fully detached component. Its effect on the optical attenuation, however, appears to be relatively modest. This is consistent with the analysis of \citet{Sanchez2026}, who examined the radial behaviour of the H$\beta$/H$\alpha$ ratio and found no significant radial variations, supporting a scenario in which the internal dust extinction is generally low.

This broad spatial correspondence is consistent with a scenario in which dust and ionised gas coexist within the same large-scale nebular structure, although the optical reddening measured here should not be interpreted as a direct tracer of the small-scale molecular knots. Previous studies of the Helix Nebula based on high-resolution imaging and spectroscopic observations have revealed a disk-like or toroidal structure surrounding the central cavity \citep{Meaburn1998, ODell1998, ODell2004, Meaburn2005}. At HST resolution, individual molecular knots can produce strong local extinction signatures. However, these knots are much smaller than the LVM fibre footprint, and their effect on the measured Balmer decrement is therefore expected to be strongly diluted. Moreover, if the knots are optically thick at optical wavelengths, their contribution would mainly depend on their area covering factor and would produce largely grey attenuation rather than a strong reddening signal.

Our extinction map is therefore consistent with a picture in which dust is present in the clump outer ring or toroidal structure, but produces only modest wavelength-dependent attenuation of the ionised gas at the spatial resolution of the LVM data. Thus, the extinction map does not by itself provide evidence of strong localised reddening associated with the torus; instead, it indicates that any internal dust extinction averaged over the LVM spaxels is generally low.

\section{Electron density of the ionised gas}
\label{sec:density_determinations}

The depth and spectral coverage of the LVM data have enabled the detection of multiple emission lines suitable for deriving a variety of $T_{\rm e}$ and $n_{\rm e}$ diagnostics, providing complementary and detailed information on the gaseous structure of the nebula.

\begin{figure}
\centering
\includegraphics[width=1.0\linewidth]{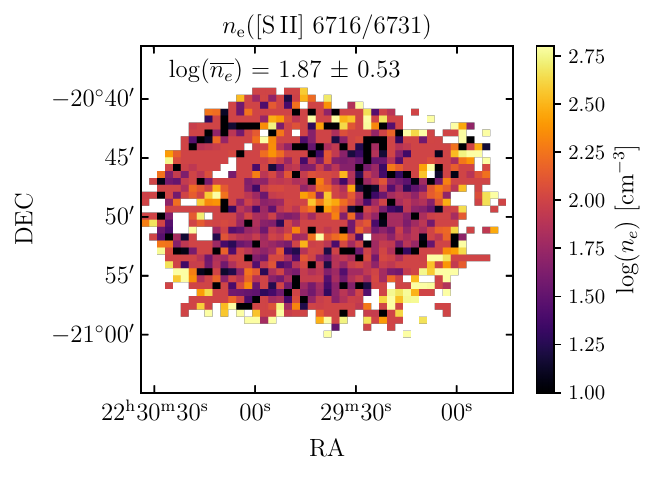}
\caption{Spatial distribution of the electron density in the Helix Nebula derived from the \sii~$\lambda6716/\lambda6731$ ratio. The map shows the median value of the electron density distribution, computed from the distribution of individual spaxel measurements. This diagnostic traces the low-ionisation regions, highlighting, within the line of sight, the denser material associated with the nebular shell and outer structures.}
\label{fig:ne_SII}
\end{figure}

\begin{figure}
\centering
\includegraphics[width=1.0\linewidth]{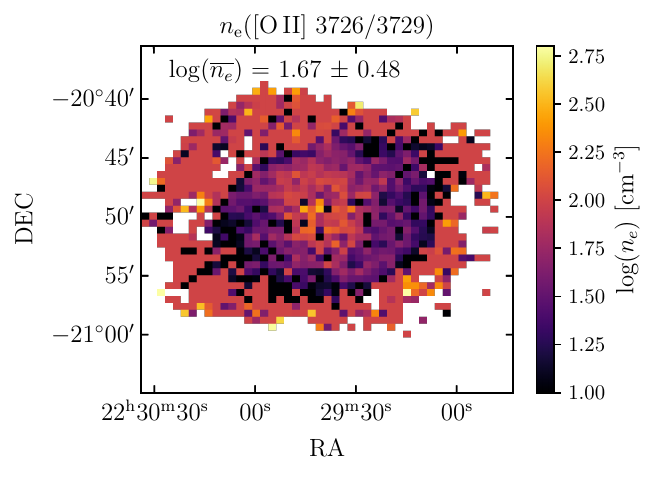}
\caption{Spatial distribution of the electron density in the Helix Nebula derived from the \oii~$\lambda3726/\lambda3729$ ratio. The map shows the median value of the electron density distribution, computed from the distribution of individual spaxel measurements. This tracer is sensitive to intermediate-ionisation regions within the line of sight, providing a complementary view of the density structure across the nebula.}
\label{fig:ne_OII}
\end{figure}

\begin{figure}
\centering
\includegraphics[width=1.0\linewidth]{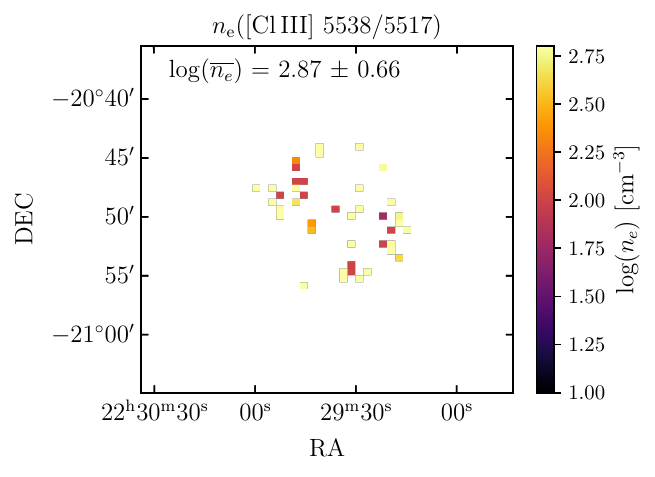}
\caption{Spatial distribution of the electron density in the Helix Nebula derived from the \cliii~$\lambda5517/\lambda5538$ ratio. The map shows the median value of the electron density distribution, computed from the distribution of individual spaxel measurements. This diagnostic probes higher-ionisation regions, revealing the density distribution closer to the central ionised cavity in the line of sight.}
\label{fig:ne_ClIII}
\end{figure}

Figures~\ref{fig:ne_SII}, \ref{fig:ne_OII}, and \ref{fig:ne_ClIII} present the $n_{\rm e}$ maps derived from the \sii~$\lambda6716/\lambda6731$, \oii~$\lambda3726/\lambda3729$, and \cliii~$\lambda5517/\lambda5538$ diagnostics, respectively. The first two diagnostics primarily trace the low-ionisation regions, while the \cliii~ diagnostic probes the intermediate-ionisation zone, as Cl$^{2+}$ has an ionisation potential range similar to that of S$^{2+}$, making it sensitive to regions closer to the ionisation front. However, in the case of the chlorine diagnostic, reliable solutions are obtained only for a limited number of spaxels because of the difficulty in detecting both lines with sufficient signal-to-noise ratio. Adaptive spatial binning or smoothing could in principle increase the S/N of the \cliii~diagnostic and provide more complete spatial coverage. However, we do not apply such a procedure here in order to preserve the native spatial resolution of the LVM data and maintain a homogeneous spaxel-by-spaxel analysis for all diagnostics. A dedicated treatment of adaptive binning would require assessing its impact separately for each line and diagnostic, and is therefore left for future work.

Several studies \citep{Peimbert:1971,Rubin:89,mendezDelgado2023b,MendezDelgado:2024,Rickards:24} have shown that the densities inferred from these diagnostics are strongly influenced by their respective sensitivity ranges, often more so than by the actual ionic stratification of the gas, particularly in low-density environments. In particular, diagnostics such as \cliii~$\lambda5517/\lambda5538$ may yield systematically higher densities than \oii~$\lambda3726/\lambda3729$ and \sii~$\lambda6716/\lambda6731$, as their sensitivity regime typically begins at $n_{\rm e} \sim 10^{3}\ \mathrm{cm}^{-3}$, whereas the latter are sensitive to densities approximately an order of magnitude lower.

In agreement with this behaviour, our maps indicate that most of the nebula lies within the low-density regime for all three diagnostics. The \sii~ and \oii~ ratios consistently place the bulk of the emission at or near their low-density limits, with only localised enhancements primarily toward the inner regions of the nebula showing moderate increases in $n_{\rm e}$ up to $\sim 200\ \mathrm{cm}^{-3}$. These enhancements likely reflect variations in path length and local density contrasts within the nebular shell.

\begin{figure*}
\centering
\includegraphics[width=1.0\linewidth]{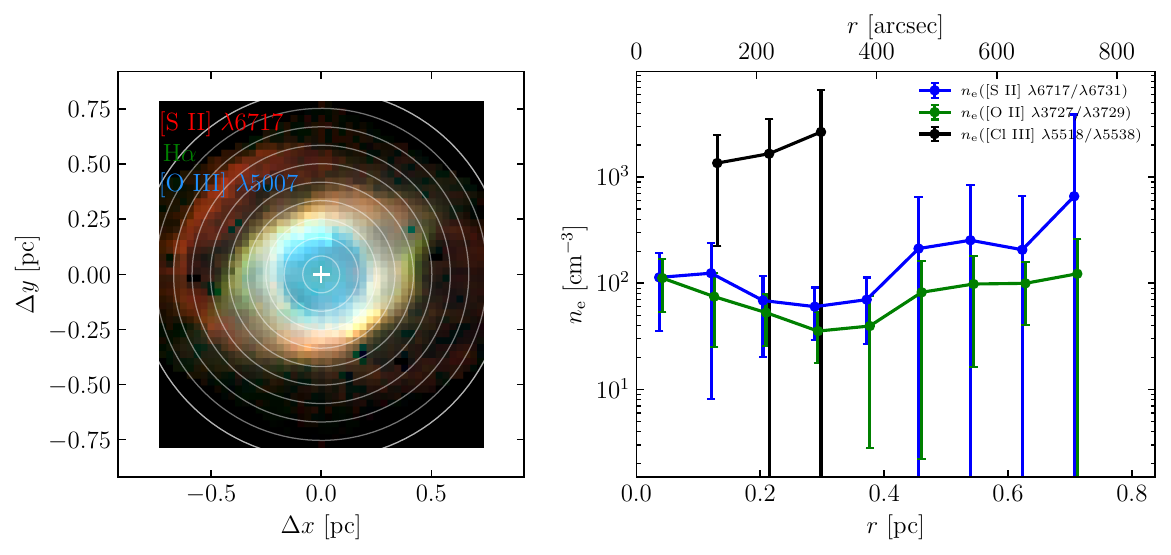}
\caption{(Left): RGB composite of the Helix Nebula. The red, green, and blue channels correspond to \sii~$\lambda6717$, H$\alpha$, and \oiii~$\lambda5007$, respectively. The white circles display the radial bins used to plot the density profile in this same figure. (Right): Azimuthally averaged radial profiles of the electron density across the Helix Nebula derived from the \sii~$\lambda6716/\lambda6731$, \oii~$\lambda3726/\lambda3729$, and \cliii~$\lambda5517/\lambda5538$ diagnostics (right). The profiles were computed in concentric radial bins centred at $(\alpha,\delta)=({\rm 22h29m38.54},{\rm -20d50m13.74})$. Symbols indicate the median density within each bin, while the error bars represent the $16^{\rm th}$ and $84^{\rm th}$ percentiles of the spaxel distribution. Most bins remain within the low-density regime, with only mild enhancements toward the inner and outer nebular regions. The systematically higher values inferred from \cliii~ reflect the different sensitivity ranges of this diagnostic.}
\label{fig:Radial_ne}
\end{figure*}

Figure~\ref{fig:Radial_ne} presents the azimuthally averaged radial density profiles derived from the three diagnostics considered in this work. The same set of concentric radial bins was used for all diagnostics and is shown in the left-hand panel as white circles overlaid on the RGB composite, indicating the portions of the nebula sampled by each bin in the density profiles shown in the right-hand panel. Bins containing fewer than five valid density measurements were excluded from the profiles. This criterion explains the different number of plotted points among diagnostics, particularly for \cliii, whose density map is more sparsely sampled owing to the lower signal-to-noise of the diagnostic lines. The profiles confirm that the Helix Nebula is predominantly a low-density object, with characteristic values of $n_{\rm e}\sim10^{2}\ \mathrm{cm}^{-3}$ over most of its extent. Both the \sii~ and \oii~ diagnostics show relatively flat distributions, with a shallow minimum at intermediate radii and modest increases toward the inner nebula and outer shell, consistent with projection. This behaviour is broadly consistent with the projected shell geometry of the nebula, where longer path lengths through the emitting gas are expected near bright rings and denser structures.

However, it is important to note that spaxel-to-spaxel variations are largely dominated by observational noise, which exceeds the intrinsic variations expected within the low-density regime of these diagnostics. In this regime, density-sensitive line ratios become weakly dependent on $n_{\rm e}$, limiting their diagnostic power. Indeed, a change of a factor of fifty in density (from $1\ \mathrm{cm}^{-3}$ to $50\ \mathrm{cm}^{-3}$) produces only a $\sim$4\% variation in the \sii~$\lambda6716/\lambda6731$ ratio, comparable to the typical noise level even in strong emission lines in deep photoionised spectra. In some cases, particularly for the \oii~$\lambda3726/\lambda3729$ diagnostic, the central values of the measured ratios can fall outside the theoretical bounds set by the Einstein radiative transition coefficients, although the overall distribution remains consistent once observational uncertainties are taken into account. In some cases, the measured ratios lie very close to the low-density limit. Consequently, a fraction of the Monte Carlo realisations can fall outside the theoretical bounds of the diagnostic, yielding undefined or poorly constrained densities. In particular, for the \oii~$\lambda3726/\lambda3729$ diagnostic, this affects $\sim$18\% of the spaxels passing the S/N cut. These cases are therefore interpreted as low-density-limit measurements rather than anomalous densities. For these spaxels, we adopt a representative value of $n_{\rm e} = 100 \pm 100\ \mathrm{cm}^{-3}$, a choice that has negligible impact on the temperature and abundance determinations because these diagnostics are largely insensitive to density in the low-density regime.

While this approach may appear simplistic, it reflects the fact that, in this density regime, the optical density diagnostics are dominated by observational uncertainties as well as uncertainties in the underlying atomic data (e.g. collision strengths and radiative transition probabilities) \citep{Morisset:2020, Mendoza:2023}. This limitation prevents a more precise determination of the density structure, restricting the analysis to a broad description in which the nebula is characterised by generally low densities with modest enhancements toward the central and outer shell regions.

The \cliii~ diagnostic, when measurable, yields systematically higher densities, typically around or above $10^{3}\ \mathrm{cm}^{-3}$. However, it could be derived for only 41 out of 1772 spaxels within the observed region, corresponding to 2.3\%, and should therefore be interpreted with caution. Given its limited spatial coverage and higher sensitivity threshold, this diagnostic likely samples only the densest regions along the line of sight. As discussed above, this offset is expected from the higher sensitivity regime of the \cliii~$\lambda5517/\lambda5538$ diagnostic rather than necessarily implying a physically distinct dense component. Overall, the radial profiles reinforce the view that the Helix Nebula is characterised by low densities with only moderate spatial variations, in agreement with previous studies such as \citet{ODell1998}.

\section{Electron temperature of the ionised gas}
\label{sec:temperature_determinations}

\begin{figure}
\centering
\includegraphics[width=1.0\linewidth]{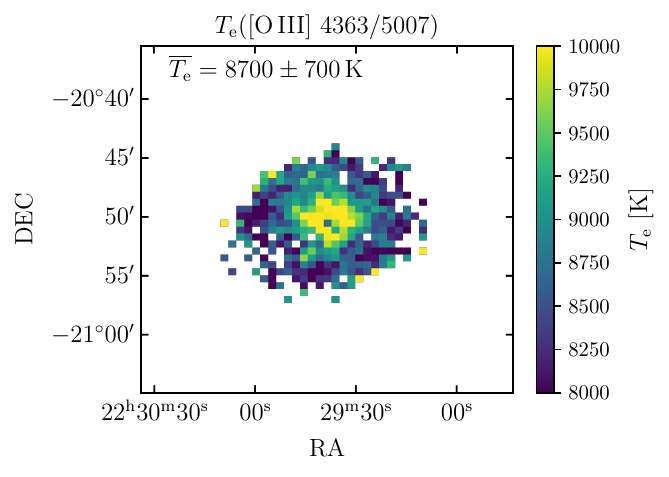}
\caption{Spatial distribution of the electron temperature in the Helix Nebula derived from the \oiii~$\lambda4363/(\lambda4959+\lambda5007)$ ratio. The map shows the median value of the electron temperature distribution, computed from the distribution of individual spaxel measurements. This diagnostic probes high-ionisation regions, tracing the temperature structure closer to the central ionised cavity along the line of sight.}
\label{fig:Te_OIII}
\end{figure}

\begin{figure}
\centering
\includegraphics[width=1.0\linewidth]{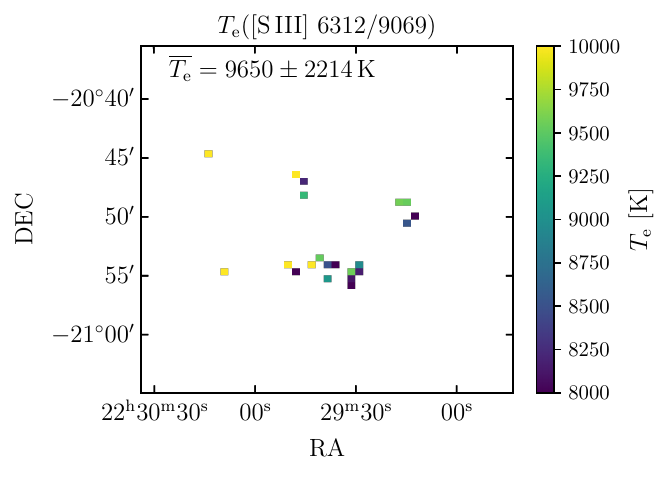}
\caption{Spatial distribution of the electron temperature in the Helix Nebula derived from the \siii~$\lambda6312/(\lambda9069+\lambda9531)$ ratio.  The map shows the median value of the electron temperature distribution, computed from the distribution of individual spaxel measurements. This diagnostic probes intermediate-ionisation regions, providing a view of the thermal structure within the main ionised shell. Only a few detections are available, mainly limited by the faintness of \siii~$\lambda 6312$.}
\label{fig:Te_SIII}
\end{figure}

\begin{figure}
\centering
\includegraphics[width=1.0\linewidth]{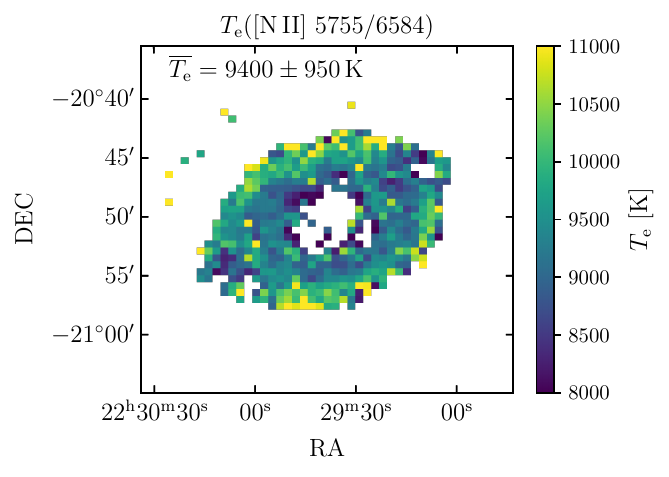}
\caption{Spatial distribution of the electron temperature in the Helix Nebula derived from the \nii~$\lambda5755/(\lambda6548+\lambda6583)$ ratio. The map shows the median value of the electron temperature distribution, computed from the distribution of individual spaxel measurements. This diagnostic traces low-ionisation regions, highlighting the thermal conditions in the outer nebular layers along the line of sight.}
\label{fig:Te_NII}
\end{figure}

\begin{figure}
\centering
\includegraphics[width=1.0\linewidth]{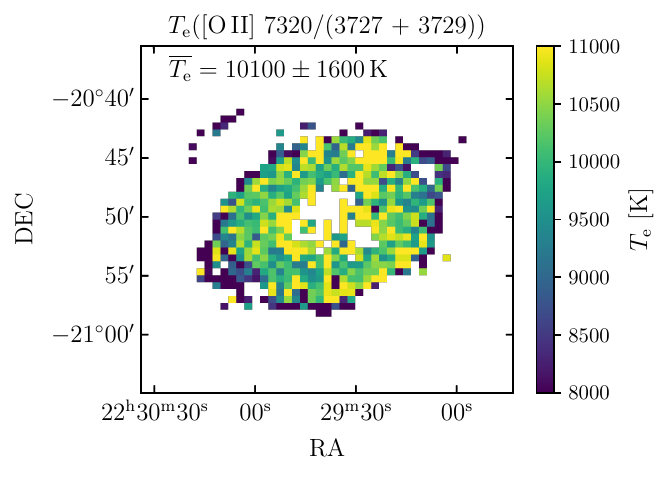}
\caption{Spatial distribution of the electron temperature in the Helix Nebula derived from the \oii~$(\lambda7319+\lambda7320)/(\lambda3726+\lambda3729)$ ratio. The map shows the median value of the electron temperature distribution, computed from the distribution of individual spaxel measurements. This diagnostic is sensitive to low-ionisation regions, providing complementary constraints on the thermal structure across the nebula along the line of sight.}
\label{fig:Te_OII}
\end{figure}

\begin{figure}
\centering
\includegraphics[width=1.0\linewidth]{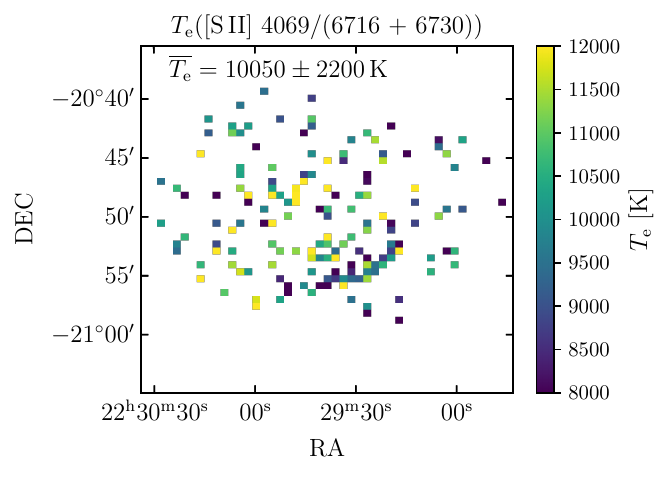}
\caption{Spatial distribution of the electron temperature in the Helix Nebula derived from the \sii~$\lambda4069/(\lambda6716+\lambda6731)$ ratio. The map shows the median value of the electron temperature distribution, computed from the distribution of individual spaxel measurements. This diagnostic traces low-ionisation regions, although it is more affected by noise and provides weaker constraints on the temperature structure.}
\label{fig:Te_SII}
\end{figure}

In Figs.~\ref{fig:Te_OIII}--\ref{fig:Te_SII}, we present the maps corresponding to the different electron temperature diagnostics discussed above. The fact that the nebula lies predominantly in the low-density regime allows the temperature diagnostics described in Sect.~\ref{sec:determining_physical_conditions} to be largely insensitive to $n_{\rm e}$, with no significant degeneracies with this parameter. 

The maps derived from the different temperature diagnostics exhibit varying spatial coverage across the field of view, primarily due to differences in the intrinsic brightness of the emission lines involved, particularly the auroral transitions. The most limited diagnostic is $T_{\rm e}(\text{\siii}~\lambda6312/\lambda9069)$, shown in Fig.~\ref{fig:Te_SIII}, owing to the difficulty of detecting the \siii~$\lambda6312$ line with sufficient signal-to-noise ratio in a significant number of spaxels.

This ion probes the intermediate-ionisation zone, where S$^{2+}$ is typically the dominant ionisation stage relative to the total sulfur abundance. In addition, the upper level responsible for the auroral transition has a relatively low excitation energy ($\sim 3.36$ eV), which generally makes the detection of \siii~$\lambda6312$ relatively common in star-forming regions. This contrasts with other auroral lines, such as \nii~$\lambda5755$ in low-metallicity environments \citep{Mendez-Delgado2023,mendezDelgado2023b,Zinchenko:2026} or \oiii~$\lambda4363$ in more metal-rich systems \citep{Esteban:2018,ArellanoCordova:2021}, which can be significantly more challenging to detect.

However, planetary nebulae generally exhibit a well-known abundance anomaly in sulfur relative to other $\alpha$-elements. The so-called ``sulfur anomaly'' \citep{Henry:2004,Shingles:2013,Tan:2024} refers to the systematic underabundance of sulfur compared to expectations based on other $\alpha$-elements, particularly oxygen \citep{Kobayashi:2006,Nomoto:2013,Esteban:2025}. In the case of the Helix Nebula, the presence of this anomaly may help explain the low detection rate of \siii~$\lambda6312$, as the flux of this line scales approximately linearly with the abundance of S$^{2+}$. This will be discussed further in Sec.~\ref{sec:chemical_abundances}. The low detection rate of $T_{\rm e}(\text{\sii}~\lambda4069/(\lambda6716+\lambda6731))$, shown in Fig.~\ref{fig:Te_SII} is driven by the same effect, with the additional contribution that the relatively high ionisation state in the central regions leads to a low S$^{+}$/S fraction.

The maps of $T_{\rm e}(\text{\oiii}~\lambda4363/\lambda5007)$, $T_{\rm e}(\text{\nii}~\lambda5755/\lambda6584)$, and $T_{\rm e}(\text{\oii}~\lambda7320/(\lambda3726+\lambda3729))$, shown in the figures \ref{fig:Te_OIII}, \ref{fig:Te_NII} and \ref{fig:Te_OII}, respectively, provide sufficiently broad spatial coverage to trace in detail the temperature distribution of high- (first) and low-ionisation (latter two) species across the plane of the sky. Interestingly, the maps of $T_{\rm e}(\text{\oiii}~\lambda4363/\lambda5007)$ and those of $T_{\rm e}(\text{\nii}~\lambda5755/\lambda6584)$ and $T_{\rm e}(\text{\oii}~\lambda7320/(\lambda3726+\lambda3729))$ are largely complementary in their spatial distribution. The former is primarily concentrated toward the central regions of the nebula, whereas the latter exhibit limited coverage in the centre and are predominantly detected in the outer regions.

This behaviour is naturally explained by the ionisation structure of the nebula, in which the central regions are dominated by highly ionised species, while the ionisation degree decreases sharply toward the periphery, in agreement with Fig.~\ref{fig:helix_iostructure}. \textit{Nevertheless, it is important to note that these maps represent two-dimensional projections of an intrinsically three-dimensional structure.} As a result, even in the central regions, there may be non-negligible contributions from low-ionisation species along the line of sight.

The results indicate that the electron temperature is not homogeneous, but instead exhibits both radial and azimuthal variations of several thousand Kelvin. The $T_{\rm e}(\text{\oiii}~\lambda4363/\lambda5007)$ diagnostic shows a clear increase in temperature toward the central regions, reaching values of up to $\sim10,000$ K, while at intermediate radii, around $0.3$ pc from the centre ($\sim300$ arcsec), the temperature decreases to $\sim8500$ K. In the outermost regions, the temperature rises again to nearly $\sim9500$ K, producing a characteristic U-shaped radial profile.

The map of $T_{\rm e}(\text{\oii}~\lambda7320/(\lambda3726+\lambda3729))$ follows a similar trend toward the centre of the nebula, reaching temperatures close to $\sim10,000$ K, and then showing a sustained decrease outward to approximately $\sim8700$ K. By contrast, the $T_{\rm e}(\text{\nii}~\lambda5755/\lambda6584)$ diagnostic reaches its minimum values in the innermost regions, with temperatures near $\sim8700$ K, and increases steadily toward the outer nebula, attaining values close to $\sim9500$ K in the most external zones.

It must be taken into account that, under certain physical conditions, the population of the upper levels giving rise to auroral lines may result from a competition between collisional excitation and recombination processes, making their interpretation more complex \citep{Rubin:1986}. The so-called recombination contribution to auroral lines is therefore a relevant effect when interpreting temperature maps, since it may lead to overestimated $T_{\rm e}$ values if the lines are assumed to arise purely from collisional excitation. A significant recombination contribution is generally expected in metal-rich systems \citep{Stasinska:2005,Garcia-Rojas:2016,GomezLlanos:2020b}, where the enhanced cooling efficiency lowers the equilibrium temperature of the gas, favouring recombination emission over collisional excitation, whose emissivity depends exponentially on $T_{\rm e}$. In addition, for recombination to contribute to the auroral emission of an ion X$^{+i}$, a substantial population of the next ionisation stage X$^{+(i+1)}$ must be present, since the intensity of the recombination component scales approximately linearly with the abundance of that higher ionisation stage \citep{Osterbrock:2006}.

Such conditions may, in principle, be present in the Helix Nebula for several of the temperature diagnostics explored here, including $T_{\rm e}(\text{\oiii}~\lambda4363/\lambda5007)$. In this case, a non-negligible population of O$^{3+}$ would be required, which may indeed be plausible given the presence of strong \heii~emission in the central regions of the nebula (to be discussed in Sect.~\ref{sec:chemical_abundances}).

Nevertheless, while the possible presence of recombination contributions deserves discussion, it is equally important to stress the lack of consensus regarding their treatment or its quantitative impact. Several studies have addressed this issue \citep{Rubin:1986,Liu:2000,Stasinska:2005,Yuan:2011, GomezLLanos:2020}, and different methodologies and correction formulae have been proposed. Perhaps the most widely used prescriptions are those given in equations (1)--(3) of \citet{Liu:2000}. However, their practical application suffers from a serious limitation: they require prior knowledge of the ionic abundance of the next ionisation stage (e.g. O$^{3+}$/H$^{+}$ in order to estimate the correction for $T_{\rm e}(\text{\oiii}~\lambda4363/\lambda5007)$, and analogously for other diagnostics). Such abundances are not known a priori, since empirical abundance determinations themselves require an adopted temperature. An alternative could be achieved by adopting an iterative approach, in which one seeks convergence between the resulting temperature, the inferred ionic abundances, and the estimated recombination contribution. However, in the absence of independent constraints on the relevant higher ionisation stages or on the temperature of the recombining gas, such a procedure remains model-dependent and may not yield a unique physical solution. In practice, this leaves temperature as a free parameter in the interpretation of the observed fluxes, introducing a strong degeneracy between temperature and metallicity, thereby severely limiting the practical value of a deep spectrum. In this sense, one effectively falls back into a regime analogous to the so-called strong-line methods for determining chemical abundances, whose large systematic uncertainties are well known \citep{Marino:2013,Sanchez:2014,Groves:2023,Rosales-Ortega:2026}.

A more robust alternative to correct the hypothetical recombination contribution would be to use heavy-element recombination lines (e.g. C, N, O recombination lines) to derive ionic abundances independently of temperature \citep{Peimbert1967}. However, this is not feasible in the present case, as no heavy-element recombination lines were detected in our spectra. Moreover, such approach is not entirely free from uncertainties: the effective recombination coefficients available in the literature for ions such as \oii~ and \nii~ \citep{Pequignot:1991, Fang:2011,Fang:2013} are limited to selected multiplets, and their accuracy for the transitions connected to the auroral levels has not yet been thoroughly assessed.

While we acknowledge the possible presence of recombination contributions, and even additional channels such as fluorescence or phosphorescence (Morisset et al., in prep.), we consider that there is no robust observational methodology to correct for them in the present data. In the absence of detected heavy-element recombination lines, the ionic abundances required to apply prescriptions such as those proposed by \citet{Liu:2000} cannot be independently constrained, making any correction highly uncertain. We therefore leave the diagnostics uncorrected and do not apply any correction for hypothetical recombination contributions to the $T_{\rm e}$ maps.

\begin{figure*}
\centering
\includegraphics[width=0.8\linewidth]{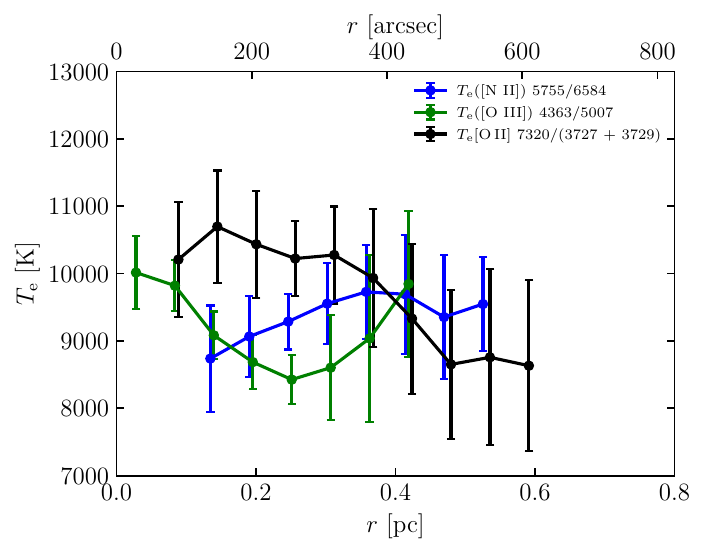}
\caption{Azimuthally averaged radial profiles of the electron temperature derived from the different diagnostics across the Helix Nebula (see Figs.~\ref{fig:Te_OIII}--\ref{fig:Te_SII}). The profiles were computed in concentric radial bins centred at $(\alpha,\delta)=({\rm 22h29m38.54},{\rm -20d50m13.74})$. The profiles reveal distinct behaviours among the ionic zones. The $T_{\rm e}(\text{\oiii}~\lambda4363/\lambda5007)$ diagnostic exhibits a characteristic U-shaped pattern, with enhanced temperatures in the central and outer regions and a minimum at intermediate radii. In contrast, $T_{\rm e}(\text{\nii}~\lambda5755/\lambda6584)$ increases outward, while $T_{\rm e}(\text{\oii}~\lambda7320/(\lambda3726+\lambda3729))$ decreases with radius and approaches the \nii-based values in the external nebula. These trends are consistent with the combined effects of central photoionisation heating, possible mechanical heating in the outer shell, and ionisation stratification along the line of sight.}
\label{fig:Radial_Te}
\end{figure*}

Nevertheless, we recognise that some diagnostics, particularly $T_{\rm e}(\text{\oii}~\lambda7320/(\lambda3726+\lambda3729))$, display behaviour that may be qualitatively consistent with the presence of recombination contributions. In principle, diagnostics such as $T_{\rm e}(\text{\oii}~\lambda7320/(\lambda3726+\lambda3729))$ and $T_{\rm e}(\text{\nii}~\lambda5755/\lambda6584)$ are both expected to probe the low-ionisation zones of the nebula and should therefore show broad consistency not only in their spatial distributions but also in their absolute values. Although photoionisation models predict modest offsets between them, such differences are generally expected to remain relatively small under normal nebular conditions \citep{ValeAsari2016}. However, the azimuthally averaged radial profiles shown in Fig.~\ref{fig:Radial_Te} indicate that, in the central regions, $T_{\rm e}(\text{\oii}~\lambda7320/(\lambda3726+\lambda3729))$ follows a trend more similar to that of $T_{\rm e}(\text{\oiii}~\lambda4363/\lambda5007)$ than to $T_{\rm e}(\text{\nii}~\lambda5755/\lambda6584)$. This behaviour could be naturally explained if the \oii~$\lambda7320,\lambda7330$ auroral lines receive a non-negligible recombination contribution from O$^{2+}$, whose abundance is highest in the central high-ionisation regions.

Although a recombination contribution to \nii~$\lambda5755$ may also be present, it is generally expected to be significantly smaller than for the \oii~auroral lines (see the discussion in Sect.~6.1.2 of \citet{mendezDelgado2023b} and particularly their Fig.~6). Therefore, while we do not apply explicit corrections, the behaviour of the \oii-based diagnostic suggests that recombination effects may contribute to part of the observed discrepancies among the low-ionisation temperature indicators.

Toward the centre, Fig.~\ref{fig:Radial_Te} shows that the enhancement of $T_{\rm e}(\text{\oiii}~\lambda4363/\lambda5007)$ can be attributed to the stronger energy input from the ionising central star, likely dominated by radiative heating. In contrast, the outward increase observed in $T_{\rm e}(\text{\nii}~\lambda5755/\lambda6584)$ may be linked to the dynamical state of the low-ionisation gas, where the expansion of the nebula and its interaction with the surrounding medium can provide an additional source of mechanical energy \citep{Peimbert:91,Draine:93}. However, this interpretation should be treated with caution, since the observed kinematics do not uniquely trace turbulent heating or shocks.

The Helix is known to be a highly evolved nebula with a roughly Hubble-type velocity pattern \citep{Meaburn2008}. In this scenario, the large-scale velocity field can arise naturally from the projection of multiple expanding shells or bipolar components with inclined axes, as discussed in previous studies \citep{Meaburn1998,Meaburn2005,ODell2005,Meaburn2008}. More recently, \citet{Sanchez2026} reported a similar behaviour using LVM H$\alpha$ kinematical data from a single exposure, interpreting the large-scale redshifted--blueshifted pattern as the projected signature of a slowly expanding, moderately inclined shell.

\begin{figure}
\centering
\includegraphics[width=1.0\linewidth]{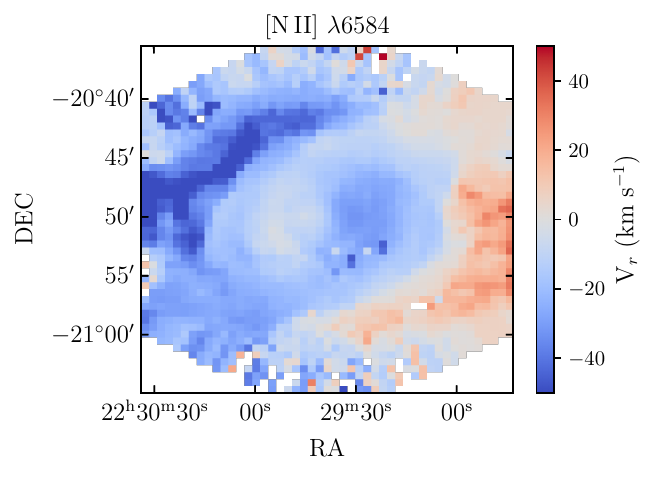}\\
\includegraphics[width=1.0\linewidth]{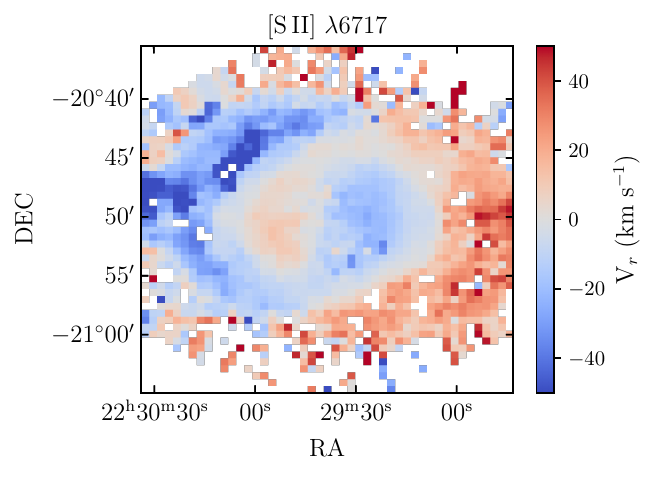}
\caption{Spatial distribution of the heliocentric line-of-sight radial velocity, ($V_{r}$), derived from the centroids of selected emission lines across the Helix Nebula. Top: radial velocity map obtained from the \nii~($\lambda6584$) emission line. Bottom: radial velocity map obtained from the \sii~($\lambda6716$) emission line. Both maps reveal increasing absolute velocities toward the outer regions, consistent with the projected expansion pattern of the nebular shell.}
\label{fig:radial_VelNII}
\end{figure}

Beyond characterising the expansion of the Helix, the spatially resolved velocity field provides an additional link between the ionisation morphology and the dynamical structure of the nebula. Indeed, Fig.~\ref{fig:radial_VelNII} presents the line-of-sight gas velocity derived from the \nii~$\lambda6584$ line. The observed radial behaviour and the two-component velocity pattern are broadly consistent with the presence of a Hubble-type flow, in which the expansion velocity increases with distance from the central star. This behaviour likely traces a combination of shell expansion, unresolved kinematic components, and line-of-sight projection effects associated with the inclined bipolar or shell-like geometry of the Helix. The \sii~velocity map show in the bottom panel provides additional information regarding the low-ionisation structures seen in the emission line map (Fig.~\ref{fig:helix_iostructure}). In particular, the prominent \sii~-bright arc toward the north-western side of the nebula is predominantly blueshifted, consistent with its association with the approaching side of the expanding shell and therefore with foreground low-ionisation component. However, the presence of redshifted \sii~emission toward the eastern side indicates that the \sii~emitting gas is not confined to a single foreground layer, but instead traces a projected, three-dimensional shell structure. These motions indicate that the outer low-ionisation gas is dynamically more complex and may be affected by weak shocks or shell--ambient medium interaction, but the present data do not allow us to isolate these contributions unambiguously. Thus, while mechanical energy may contribute to the thermal structure of the low-ionisation gas, the observed velocity field should be interpreted primarily as a signature of the three-dimensional expansion geometry of the Helix, rather than as direct evidence for turbulence-driven heating.

The presence of this non-uniform temperature structure provides new insight into the so-called ``abundance discrepancy problem'', namely the systematic difference between abundances derived from CELs and those obtained from RLs of heavy elements. Since \citet{Peimbert1967}, it has been proposed that temperature inhomogeneities in photoionised nebulae may be responsible for this discrepancy, owing to the strong exponential dependence of CEL emissivities on $T_{\rm e}$. Even if part of the observed $T_{\rm e}$ variations were instead produced by other effects, such as the aforementioned recombination contributions, the resulting systematic bias on CEL-based abundances would be qualitatively similar.

Although we do not detect heavy-element RLs in the datacube constructed from the current set of observations, the observed temperature structure is consistent with the presence of such inhomogeneities, as suggested by recent observational studies \citep{Mendez-Delgado2023}. However, while strong temperature variations have indeed been reported in planetary nebulae, in agreement with the results presented here, their physical origin may instead be linked to the presence of chemically inhomogeneous, metal-rich clumps, which can locally enhance the cooling efficiency of the gas and generate apparent temperature fluctuations. It has also been suggested that RLs and CELs may arise from different regions due to morphological and structural differences associated with the three-dimensional ejection and distribution of the nebular material. \citep{Torres-Peimbert:1990,Liu:2000,GomezLLanos:2020,GarciaRojas:22,Ruiz-Escobedo:2022,Richer-etal:2022}.

\section{Chemical abundances of the ionised gas}
\label{sec:chemical_abundances}

Once the electron density and temperature structure of the gas had been established, the extinction-corrected emission-line fluxes were used to derive ionic abundances for the different species detected in the nebula. Owing to the natural ionisation stratification of the gas, an appropriate temperature scheme must be adopted, such that low-ionisation temperature diagnostics are used for ions arising predominantly in low-ionisation zones, while high-ionisation diagnostics are applied to ions formed in more highly ionised regions \citep{OrteGarcia:2026}.

Following the standard prescriptions of \citet{mendezDelgado2023b}, we adopted $T_{\rm e}(\text{\nii}~\lambda5755/\lambda6584)$ for low-ionisation species and $T_{\rm e}(\text{\oiii}~\lambda4363/\lambda5007)$ for high-ionisation species. However, a practical limitation arises because the spatial coverage of these temperature maps is more restricted than that of several strong emission lines commonly used for abundance determinations. As shown in Sect.~\ref{sec:temperature_determinations}, the central regions of the nebula may lack measurements of $T_{\rm e}(\text{\nii}~\lambda5755/\lambda6584)$ owing to the weak detection of \nii~$\lambda5755$ in these highly ionised zones, while still displaying strong \nii~$\lambda6584$ emission from which N$^{+}$/H$^{+}$ can be estimated.

To address this issue, and motivated by the complementary spatial behaviour of the two diagnostics, we adopted the following hybrid scheme. For low-ionisation ions, $T_{\rm e}(\text{\nii}~\lambda5755/\lambda6584)$ was used by default, while in the central regions where this diagnostic is unavailable we adopted $T_{\rm e}(\text{\oiii}~\lambda4363/\lambda5007)$ as a reasonable approximation. Conversely, for high-ionisation ions we adopted $T_{\rm e}(\text{\oiii}~\lambda4363/\lambda5007)$ by default, and used $T_{\rm e}(\text{\nii}~\lambda5755/\lambda6584)$ in the outermost regions where the former diagnostic is not available, ensuring full spatial coverage of the abundance maps.

Although this approximation is not perfect, the overall consistency between both temperature structures (Fig.~\ref{fig:Radial_Te}) suggests that any introduced biases are likely to be small. Moreover, the impact on total elemental abundances is expected to be minor. In highly ionised regions where $T_{\rm e}(\text{\nii}~\lambda5755/\lambda6584)$ is unavailable, the contribution of low-ionisation ions to the total abundance budget is generally small. For intermediate-ionisation species, no dedicated temperature diagnostic with sufficient spatial coverage was available because of the limited detection of \siii~$\lambda6312$, as discussed in Sec.~\ref{sec:temperature_determinations}. These ions were therefore treated using the high-ionisation temperature scheme. Although the relation between $T_{\rm e}(\text{\siii}~\lambda6312/\lambda9069)$ and $T_{\rm e}(\text{\oiii}~\lambda4363/\lambda5007)$ is temperature-dependent, this approximation is appropriate for high-ionisation nebulae in the temperature range of the Helix, where the two diagnostics tend to converge \citep{Rogers:2026, OrteGarcia:2026}.

\subsection{Emission-line diagnostics for ionic abundances}
The high-ionisation species analysed in this work are He$^{+}$ from \hei~$\lambda5876$, He$^{2+}$ from \heii~$\lambda4686$, O$^{2+}$ from \oiii~$\lambda\lambda4959,5007$, Ne$^{2+}$ from \neiii~$\lambda3868$, and Ar$^{3+}$ from \ariv~$\lambda4740$. Intermediate-ionisation species include Ar$^{2+}$ from \ariii~$\lambda7135$, S$^{2+}$ from \siii~$\lambda9069$, and Cl$^{2+}$ from \cliii~$\lambda\lambda5517,5538$. Low-ionisation species include S$^{+}$ from \sii~$\lambda\lambda6716,6731$ and O$^{+}$ from \oii~$\lambda\lambda3726,3729$.

All transitions were individually inspected to avoid blends with neighbouring lines from other ions and to exclude measurements affected by clear observational defects such as telluric absorption or contamination by sky emission features. In the case of \hei~$\lambda5876$, we selected the brightest triplet line not significantly affected by self-absorption effects associated with the metastable $2^{3}\mathrm{S}$ level \citep{MendezDelgado:2025}.

\subsection{Oxygen ionic abundances and total O/H}
\begin{figure*}
    \centering
    \includegraphics[width=0.5\linewidth]{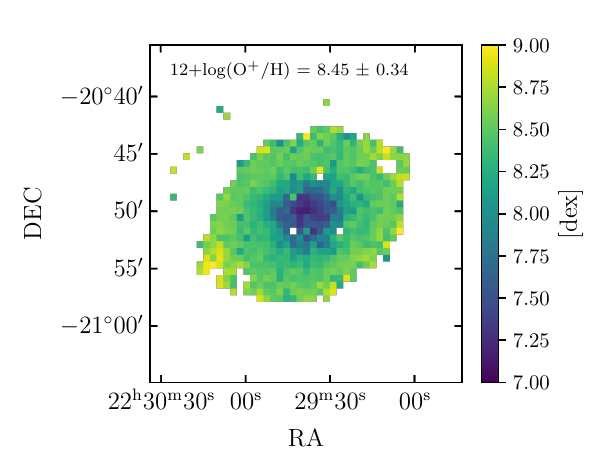}~
    \includegraphics[width=0.5\linewidth]{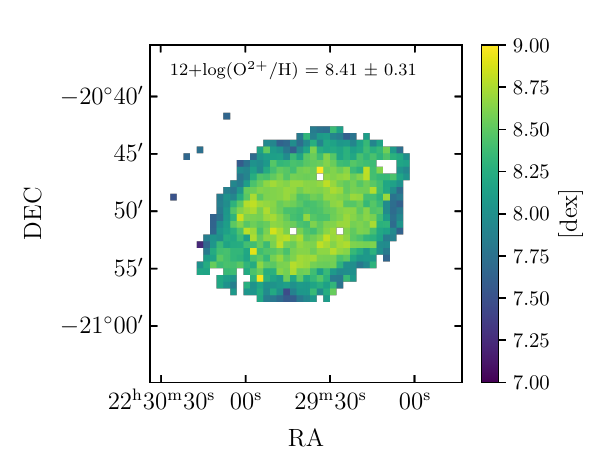}\\
    \includegraphics[width=0.5\linewidth]{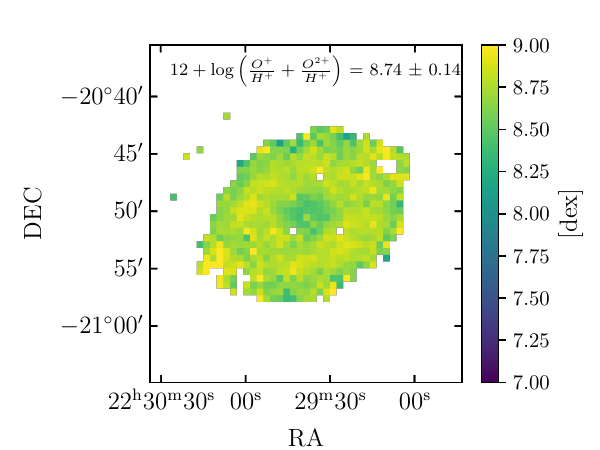}
    \label{fig:O_abundance}
    \end{figure*}

We have derived both the ionic oxygen abundances across the Helix Nebula, as described in Section~\ref{sec:determining_physical_conditions}. Figure~\ref{fig:O_abundance} presents, in the top row, the spatial distributions of O$^{+}$ and O$^{2+}$ relative to H$^{+}$, while the bottom row shows the resulting sum of the ionic abundances. In all panels, the colour scale indicates the abundance expressed as $12 + \log(\mathrm{X}/\mathrm{H})$ for the corresponding ion or for the total oxygen content.

The spatial distributions of the ionic abundances closely follow the expected ionisation structure of the nebula. The O$^{2+}$ abundance, traced primarily through the strong \oiii~$\lambda\lambda4959,5007$ lines, dominates the inner regions of the nebula and is preferentially concentrated toward the central cavity, although not in its innermost region. In contrast, O$^{+}$, traced through \oii~$\lambda\lambda3726,3729$, becomes progressively more important toward the main ring and outer shell. This behaviour is fully consistent with the classical stratified structure reported in previous studies of the Helix Nebula \citep{ODell1998,Henry1999,Meaburn2005}, and with the qualitative ionisation maps presented in Fig.~\ref{fig:helix_iostructure}. As expected for a low-density evolved planetary nebula ionised by a hot central star, the degree of ionisation decreases with increasing projected distance from the central source.

The oxygen abundance obtained from the direct sum O$^{+}$+O$^{2+}$ reveals a relatively homogeneous large-scale distribution, although with non-negligible local structure. The median value across the nebula is $12+\log(\mathrm{O/H}) = 8.73$, placing the Helix Nebula close to the Solar oxygen abundance of $12+\log(\mathrm{O/H})=8.69$ \citep{Asplund:2021}.

Near the nebular centre, the ionic sum O/H reaches values up to $\sim 0.3$ dex below the median abundance. Rather than indicating genuine chemical inhomogeneities at these spatial scales, this central ``oxygen deficit'' is best interpreted as an ionisation effect, caused by a shift of oxygen into ionisation stages not included in the direct sum. This effect is further discussed in Section~\ref{sec:oxygen_deficit}.

\subsection{Comparison with previous O/H determinations}
When compared with previous determinations, our oxygen abundance is broadly consistent with earlier studies, although the reported values span a non-negligible range depending on the adopted methodology, temperature structure, and ionisation corrections. For the Helix Nebula, the compilation of \citet{Henry1999} in their Table~5 lists values between $8.52$ and $8.95$ from previous analyses, with an average of $8.66$. Our median value of $12+\log(\mathrm{O/H})=8.73$ lies comfortably within this historical range and may be regarded as one of the most robust and internally consistent determinations currently available, owing to the homogeneous treatment of the data, the careful derivation of the physical conditions, and the wide spatial coverage of the IFU observations. In particular, the latter reduces potential systematic biases associated with the selection of specific nebular regions, some of which may contain a significant fraction of unaccounted O$^{3+}$, as discussed above.

To assess the impact of this missing ionisation stage, we computed an ICF-corrected oxygen abundance using the prescription of \cite{Delgado-Inglada:2014}. For a spatially uniform O/H abundance, a fully adequate ICF would be expected to flatten the trend with He$^{2+}$/He$^{+}$, since this ratio traces where O$^{3+}$ should become important. Although the correction increased the O/H, it was not able to fully remove the central decrease or substantially alter the spatial trend. This behaviour is not unexpected, since no ionisation correction factor can be entirely model independent \citep{Delgado-Inglada:2014, Morisset:2015}, and any quantitative recovery of O$^{3+}$ necessarily depends on assumptions about the ionisation structure of the nebula. We therefore treat the directly measured $\mathrm{O}^{+} + \mathrm{O}^{2+}$ as a robust lower limit in the central regions, where O$^{3+}$ may be significant, and as an approximately complete abundance estimate in the outer nebula, where higher ionisation stages are expected to be negligible.

\subsection{Other ionic abundances}
\begin{figure*}
    \centering
    % Helium
    \includegraphics[width=0.32\linewidth]{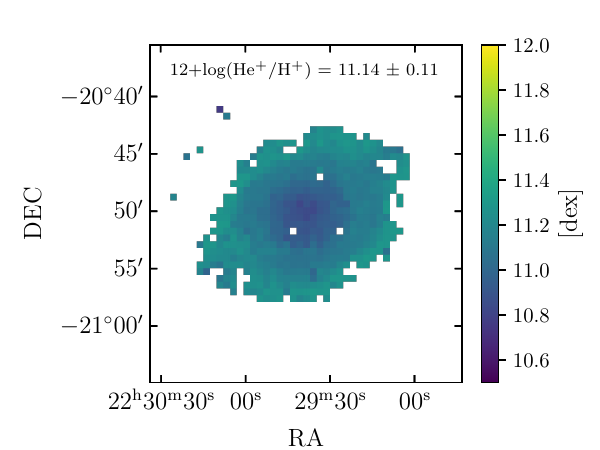}
    \includegraphics[width=0.32\linewidth]{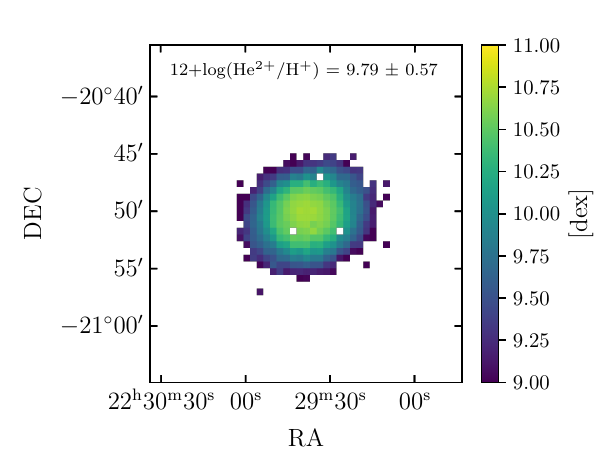}
    \includegraphics[width=0.32\linewidth]{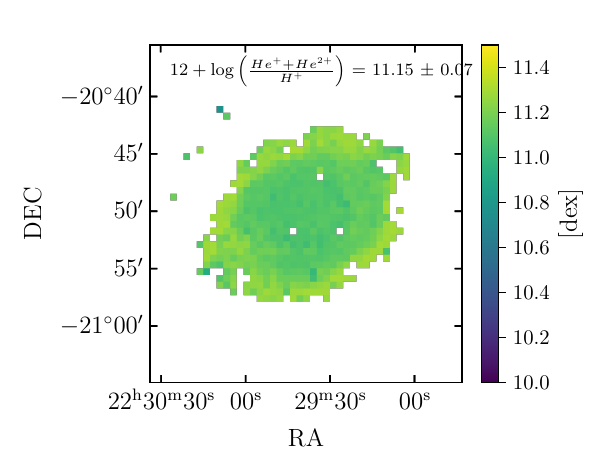}\\
    
    % Sulfur
    \includegraphics[width=0.32\linewidth]{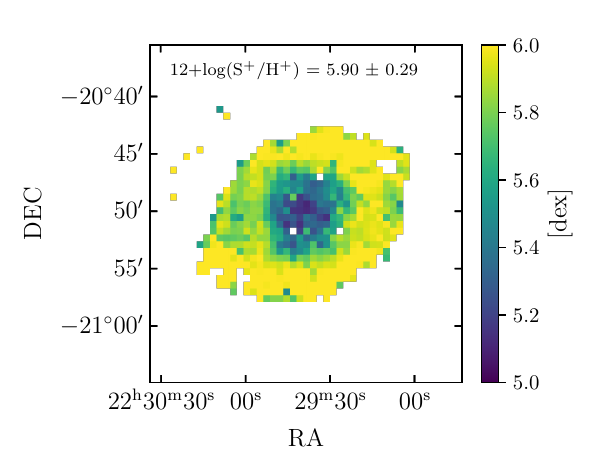}
    \includegraphics[width=0.32\linewidth]{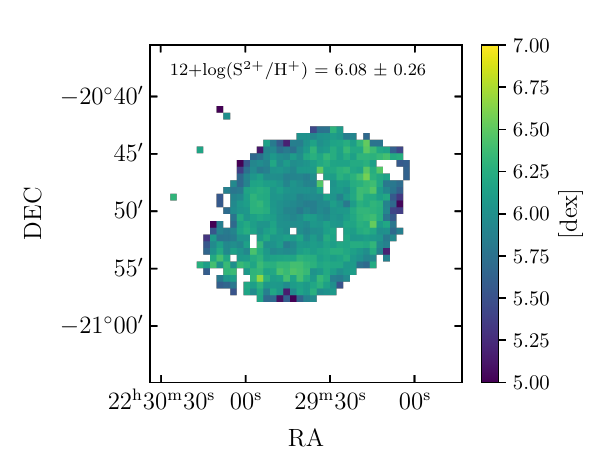}
    \includegraphics[width=0.32\linewidth]{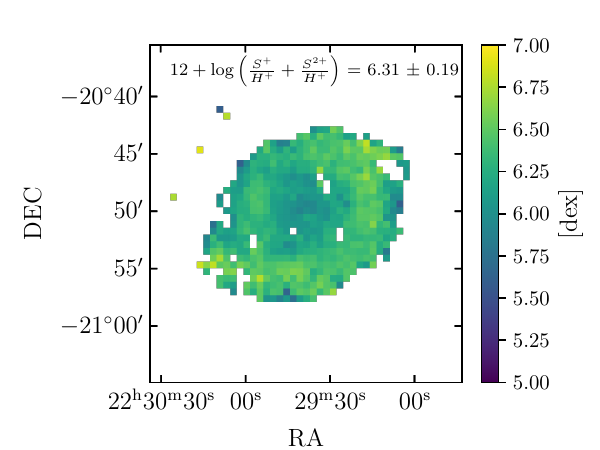}\\
    
    % Other low/intermediate/high ionisation species
    \includegraphics[width=0.32\linewidth]{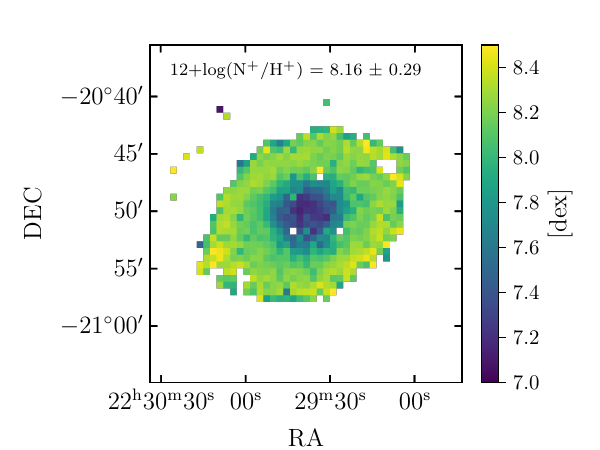}
    \includegraphics[width=0.32\linewidth]{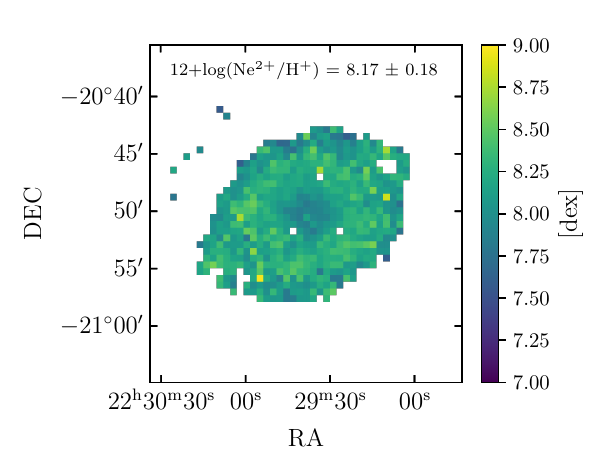}
    \includegraphics[width=0.32\linewidth]{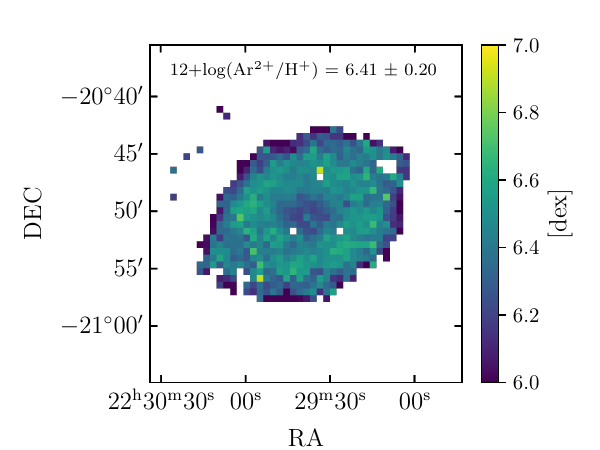}\\
    
    % Sparse/high-ionisation detections
    \includegraphics[width=0.32\linewidth]{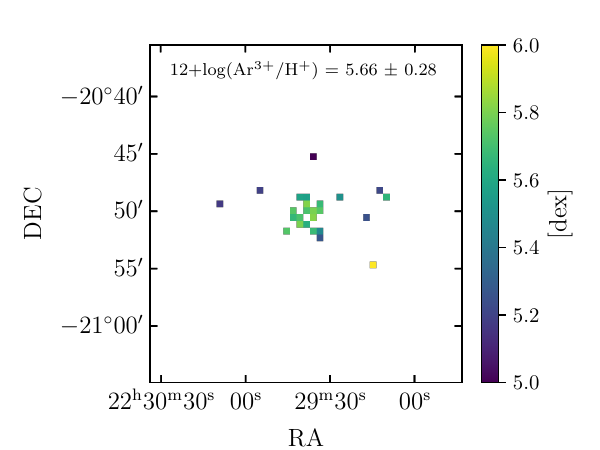}
    \includegraphics[width=0.32\linewidth]{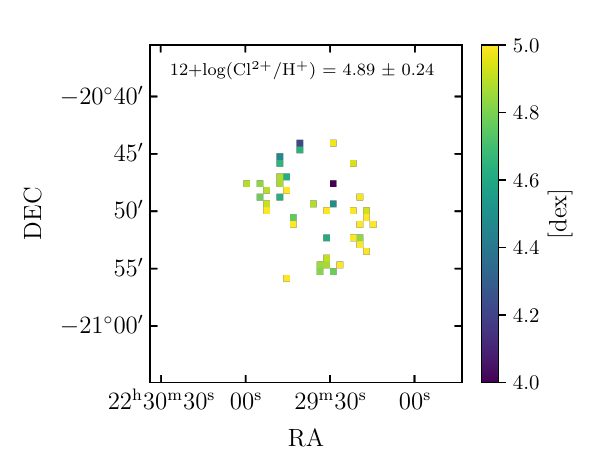}
    
    \caption{Spatial distributions of the ionic abundances derived for the main detected species in the Helix Nebula. From top to bottom: helium ions and total helium abundance (He$^{+}$, He$^{2+}$, He/H); sulfur ions and their summed abundance (S$^{+}$, S$^{2+}$, S/H); other detected species (N$^{+}$, Ne$^{2+}$, Ar$^{2+}$); and the more sparsely detected Ar$^{3+}$ and Cl$^{2+}$. All colour scales are expressed as $12+\log(\mathrm{X}/\mathrm{H})$. The distributions broadly follow the nebular ionisation structure, with higher ionisation species concentrated toward the inner regions and lower ionisation species becoming more prominent in the outer shell.}
    \label{fig:ionic_abundances_other}
\end{figure*}

The remaining ionic abundances derived in this work are presented in Fig.~\ref{fig:ionic_abundances_other}. In the cases of helium and sulfur, the abundances of their singly and doubly ionised species were summed, and the resulting maps are shown as proxies for the total elemental abundances. For helium, the sum $\mathrm{He}^{+} + \mathrm{He}^{2+}$ should closely represent the total abundance, particularly in the central regions where neutral helium is expected to be negligible. In contrast, for sulfur this sum should be regarded as a lower limit, owing to the potentially significant contribution of higher ionisation stages such as S$^{3+}$.

The spatial distributions of the different ionic abundances broadly follow the ionisation structure already described in Fig.~\ref{fig:ionic_abundances_other}. The central regions host the highest ionisation stages, including detectable He$^{2+}$ emission, while the ionisation degree decreases progressively toward the outer nebula. This behaviour is consistent with the hard radiation field produced by the hot central star and the stratified structure expected in an evolved planetary nebula.

In the case of helium, the summed abundance map reveals a relatively homogeneous distribution, with a characteristic value of $12+\log(\mathrm{He/H})\approx11.12$. This is approximately $0.2$ dex above the Solar helium abundance reported by \citet{Asplund:2021}. The derived helium enrichment is moderate, but astrophysically significant, and is naturally interpreted as the consequence of dredge-up episodes experienced by the progenitor star during its RGB and AGB evolution. During these phases, helium synthesised in the stellar interior is transported to the stellar surface and later expelled into the nebula through mass loss \citep{Karakas:2014}.

\subsection{Ionic abundance ratios and ionisation structure}
\begin{figure*}
    \centering
    % Helium / Oxygen
    \includegraphics[width=0.32\linewidth]{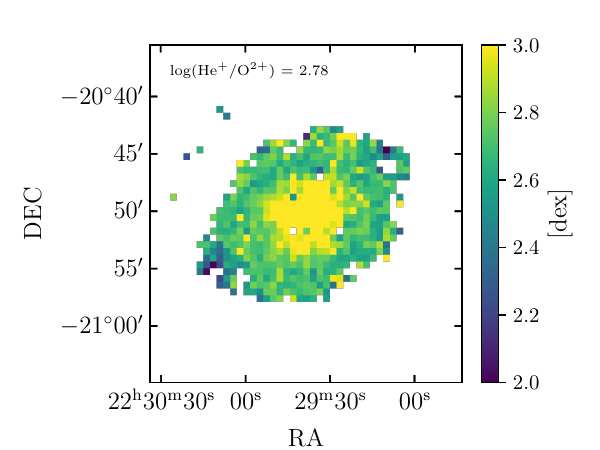}
    \includegraphics[width=0.32\linewidth]{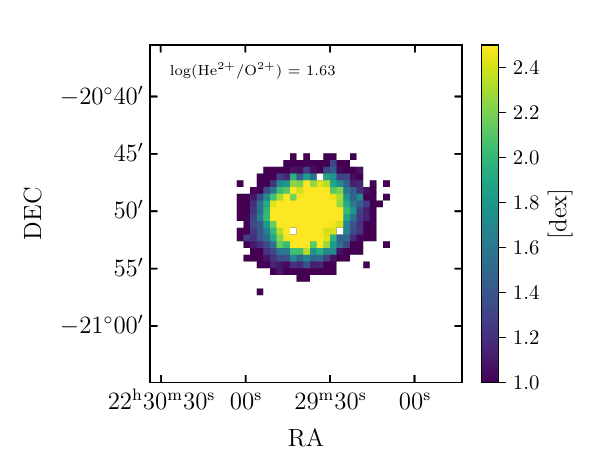}
    \includegraphics[width=0.32\linewidth]{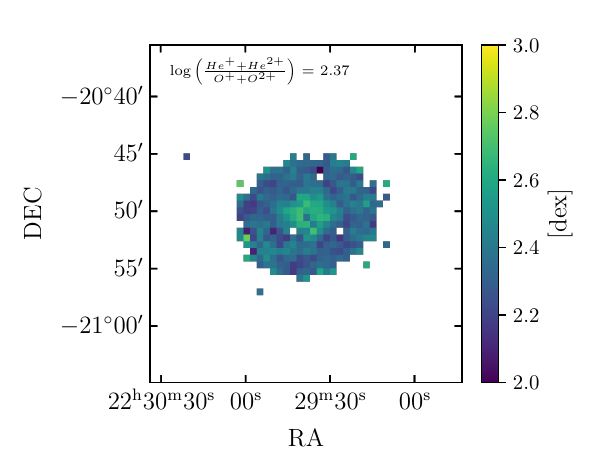}\\
    
    % Sulfur / Oxygen
    \includegraphics[width=0.32\linewidth]{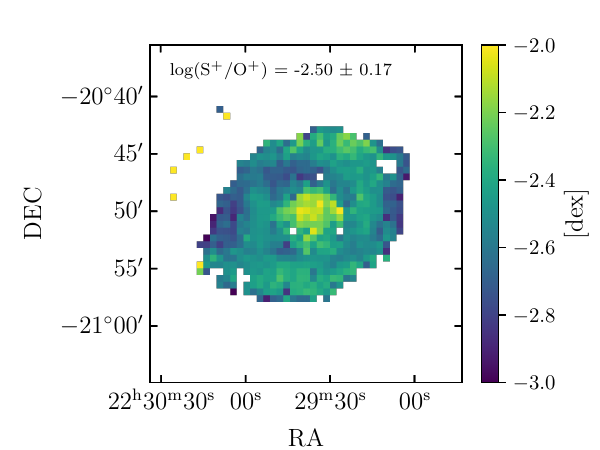}
    \includegraphics[width=0.32\linewidth]{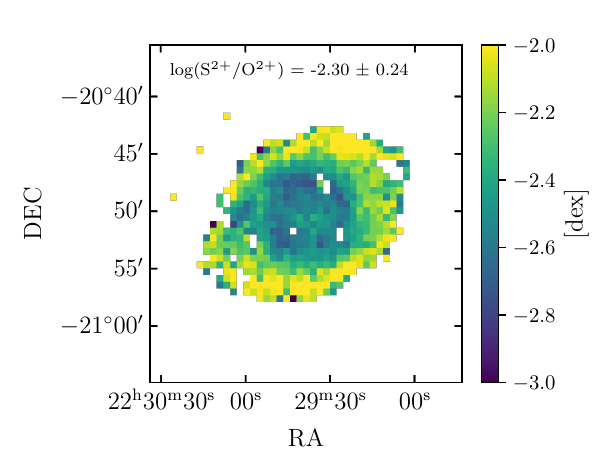}
    \includegraphics[width=0.32\linewidth]{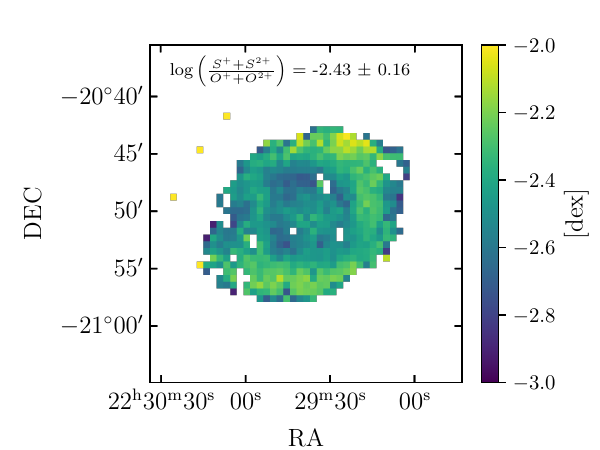}\\
    
    % Classical ionisation tracers
    \includegraphics[width=0.32\linewidth]{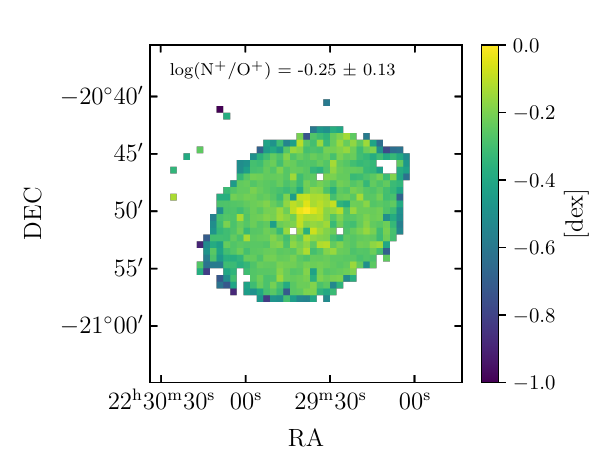}
    \includegraphics[width=0.32\linewidth]{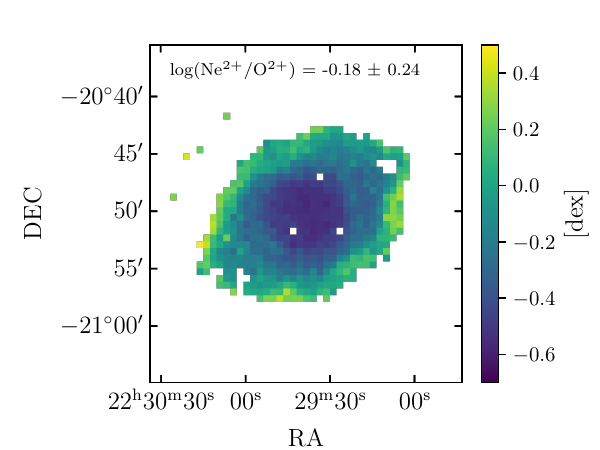}
    \includegraphics[width=0.32\linewidth]{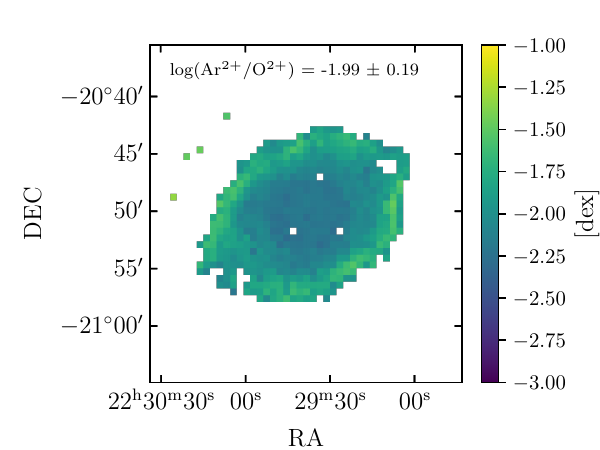}\\
    
    % Intermediate-ionisation consistency tracers
    \includegraphics[width=0.32\linewidth]{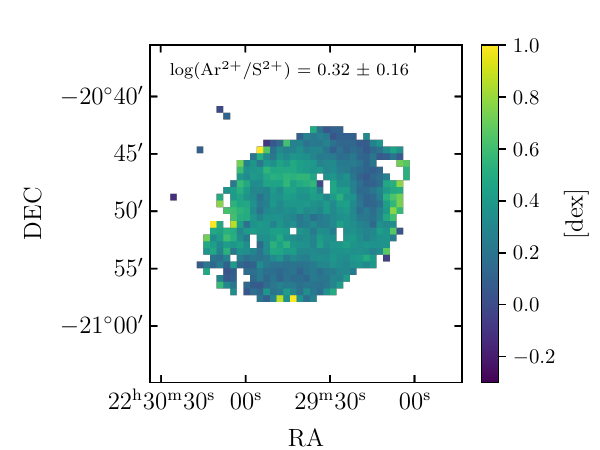}
    \includegraphics[width=0.32\linewidth]{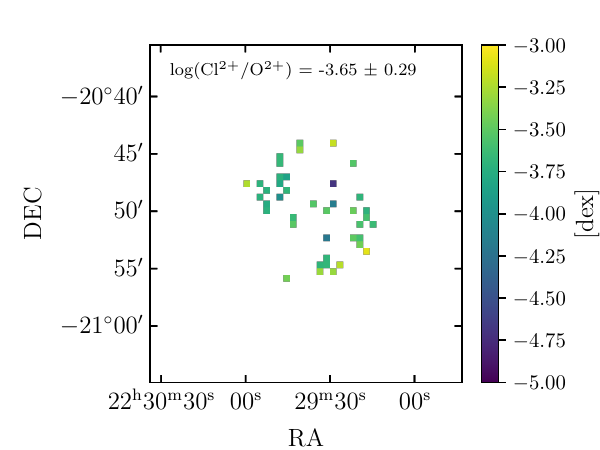}
    \includegraphics[width=0.32\linewidth]{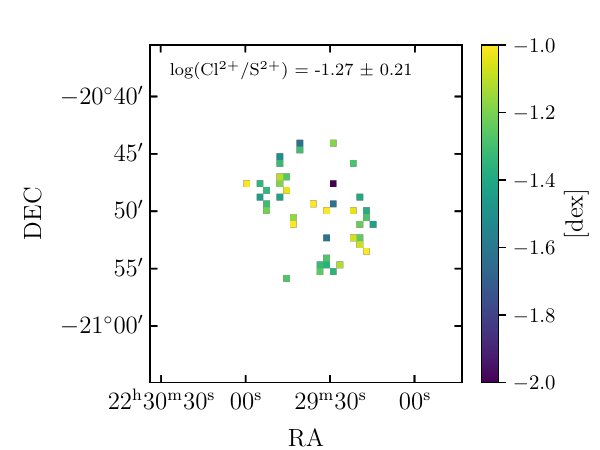}
    
    \caption{Spatial distributions of selected ionic abundance ratios across the Helix Nebula. From top to bottom: helium-to-oxygen ratios (He$^{+}$/O$^{2+}$, He$^{2+}$/O$^{2+}$, He/O), sulfur-to-oxygen ratios (S$^{+}$/O$^{+}$, S$^{2+}$/O$^{2+}$, S/O), classical ionisation tracers (N$^{+}$/O$^{+}$, Ne$^{2+}$/O$^{2+}$, Ar$^{2+}$/O$^{2+}$), and consistency ratios between ions of similar ionisation potential (Ar$^{2+}$/S$^{2+}$, Cl$^{2+}$/O$^{2+}$, Cl$^{2+}$/S$^{2+}$). All colour scales are expressed in logarithmic units. Ratios involving ions with similar ionisation structure are expected to be comparatively uniform, whereas stronger spatial gradients mainly trace variations in the local ionisation degree across the nebula.}
    \label{fig:relative_abundances}
\end{figure*}

To place the remaining ionic abundances into context and extract information beyond the pure ionisation structure, we have examined a series of ionic abundance ratios, selecting species with comparable ionisation potentials whenever possible. Figure~\ref{fig:relative_abundances} presents the resulting maps, which highlight the regions where one ionic stage dominates over another and therefore provide a complementary view of the nebular ionisation balance.

Several ratios are primarily governed by the changing hardness of the radiation field. For example, He$^{2+}$/O$^{2+}$ peaks strongly in the central regions, tracing the zones exposed to the most energetic photons, while N$^{+}$/O$^{+}$ and S$^{+}$/O$^{+}$ become comparatively enhanced in the outer shell, where lower ionisation stages dominate. These behaviours are fully consistent with the ionisation stratification discussed in previous sections.

By contrast, ratios involving ions with similar ionisation potentials are expected to be less sensitive to the local ionisation parameter and more representative of intrinsic abundance patterns and atomic physics effects. A clear example is S$^{2+}$/Ar$^{2+}$, which shows a relatively homogeneous spatial distribution across most of the nebula, indicating that both ionic stages coexist remarkably well throughout the ionised gas. 

More irregular patterns are found for ratios involving Cl$^{2+}$, mainly owing to the limited number of spaxels where the weak \cliii~lines are reliably detected. Therefore, the apparent structures in those maps should be interpreted with caution.

\begin{figure*}
    \includegraphics[width=0.8\linewidth]{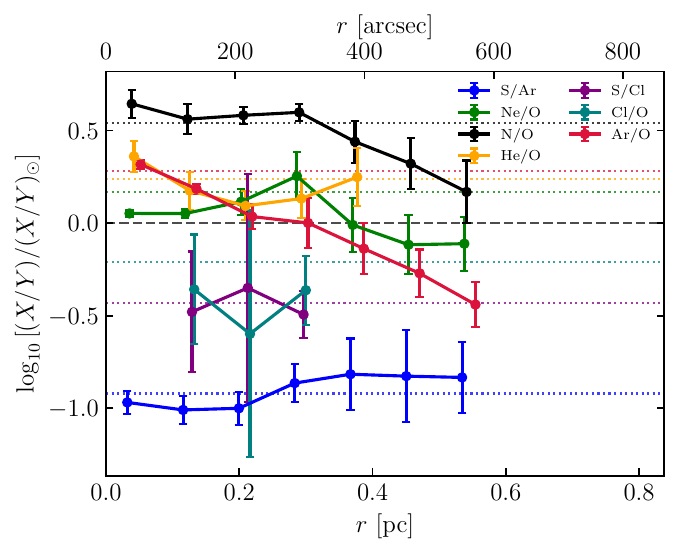}
    \caption{Azimuthally averaged radial profiles of selected ICF-corrected elemental abundance ratios, normalised to their corresponding Solar values \citep{Asplund:2021}. The vertical axis shows $\log_{10}[(X/Y)/(X/Y)_{\odot}]$, such that the black dashed horizontal line marks the Solar reference. The profiles include He/O and N/O as tracers of the chemical enrichment of the progenitor, S/Ar and S/Cl as diagnostics of the sulfur abundance, and Ne/O, Cl/O, and Ar/O as comparisons among elements whose abundances are not expected to be significantly modified during the evolution of the progenitor star. The coloured dotted horizontal lines indicate the corresponding abundance ratios derived from the integrated spectrum using the elemental abundances listed in Table~\ref{tab:integral_total_abundances}. Nitrogen is systematically enhanced relative to Solar, whereas sulfur remains underabundant with respect to Ar and Cl, consistent with the classical sulfur anomaly observed in planetary nebulae.}
    \label{fig:radial_relative_abundances}
\end{figure*}

The comparison of several ionic abundance ratios and their radial behaviour provides additional insight into the chemical composition of the nebula and into the nucleosynthetic history of its progenitor star. Figure~\ref{fig:radial_relative_abundances} shows azimuthally averaged abundance ratios normalised to their corresponding Solar values, allowing departures from the Solar abundance pattern to be directly visualised.

The abundance ratios shown in Fig.~\ref{fig:radial_relative_abundances} were corrected using ionisation correction factors (ICFs) derived from the Mexican Million Models database, 3MdB \citep{Morisset2015}. For each radial bin, we queried the grid of planetary-nebula photoionisation models and selected models whose ionisation structure matched the observed ionic ratios within a tolerance of 0.05 dex. In particular, the selection was based on the ratios He$^{2+}$/He$^{+}$, O$^{2+}$/O$^{+}$, and S$^{2+}$/S$^{+}$, whenever the three ratios were available. In the bin where the helium constraint was not available, the ICFs were instead derived using only the O$^{2+}$/O$^{+}$ and S$^{2+}$/S$^{+}$ constraints. To assess the impact of this choice, we compared the ICFs obtained from the full set of constraints, He$^{2+}$/He$^{+}$, O$^{2+}$/O$^{+}$, and S$^{2+}$/S$^{+}$, with those obtained using only the oxygen and sulfur ratios in the radial bins where all three diagnostics were available. The absolute differences between both sets of ICFs are typically below 0.1 in linear units. Therefore, to first order, the ICFs derived from the O$^{2+}$/O$^{+}$ and S$^{2+}$/S$^{+}$ constraints provide an adequate approximation for the radial bin where the helium constraint is missing. From the selected models, we computed the mean correction factors required to transform the observed ionic ratios into total elemental ratios, namely ICF(S$^{2+}$/Cl$^{2+}$), ICF(S$^{2+}$/Ar$^{2+}$), ICF(Ne$^{2+}$/O$^{2+}$), and ICF(N$^{+}$/O$^{+}$).

We emphasise that these ICFs are intended as empirical, model-guided corrections based on the ionisation structure sampled by the 3MdB grid. They do not replace tailored photoionisation modelling of the Helix Nebula, which would be required to fully account for its three-dimensional geometry, density structure, and spatially varying ionisation conditions. Therefore, the corrected abundance ratios should be interpreted with caution, as first-order estimates of the global radial trends rather than as definitive elemental abundance determinations.

Figure~\ref{fig:radial_relative_abundances} shows that the ICF-corrected N/O ratio is systematically above the Solar value over the full radial range sampled by our data. This indicates that the nitrogen enhancement persists after accounting for first-order ionisation effects through the model-guided ICFs described above, supporting a genuine enrichment of nitrogen in the Helix Nebula. The corrected profile shows a mild outward decline, from approximately 0.5--0.6 dex above Solar in the inner regions to about 0.2 dex in the outer bins. Since the ratios have already been corrected for first-order ionisation effects, this residual gradient should not be interpreted solely as a direct change in the ionic balance. Instead, it may reflect a combination of residual ionisation-structure effects not fully captured by the adopted ICFs, projection through different nebular layers, and the limitations of applying grid-based corrections to an intrinsically stratified object. Nevertheless, the fact that the corrected N/O ratio remains super-Solar throughout the nebula strongly supports nitrogen enrichment through CNO-cycle processing in the progenitor star.
For comparison, Fig.~\ref{fig:radial_relative_abundances} also shows, as coloured dotted horizontal lines, the corresponding abundance ratios derived from the integrated spectrum. These values provide a global reference against which the spatially resolved trends can be compared. In general, the integrated ratios lie within the range spanned by the radial profiles, although some differences are expected because the integrated spectrum represents an emission-weighted measurement of the nebula rather than a spatial average. The agreement is particularly clear for the sulfur-related ratios: the integrated S/Ar and S/Cl values reproduce the systematically sub-Solar levels inferred from the spatially resolved profiles. This provides additional evidence that the sulfur deficit is a global property of the Helix Nebula rather than a feature confined to particular regions.

This conclusion is reinforced by the helium abundance. The He/O ratio is also above the Solar value in both the radial profile and the value inferred from the integrated spectrum (Fig.~\ref{fig:radial_relative_abundances}). Taken together, the values of He/H and N/O place the Helix Nebula close to the classical boundary of Type~I planetary nebulae introduced by \citet{Peimbert:1978}. In particular, we obtain $12+\log(\mathrm{He/H})\approx11.12$ (He/H $\approx0.13$) together with $\log(\mathrm{N/O})\approx-0.3$, which lies near the commonly adopted threshold for this class. However, because both quantities lie only marginally above the boundary and do not indicate extreme enrichment, NGC~7293 is more appropriately described as a transitional or marginal Type~I object rather than a prototypical member of the class.

\section{Integrated-spectrum analysis}
\label{sec:integrated_spectrum}
For the analysis of the integrated spectrum, we follow a methodology identical to that described in Section~\ref{sec:spectral_analysis}, deriving the extinction, electron density, electron temperature, and ionic abundances from the same set of emission lines used for resolved maps. The main difference lies in the fact that, in the integrated spectrum, we obtain a reliable determination of $T_{\rm e}(\text{\siii}~\lambda6312/\lambda9069)$, which we adopt to estimate the abundances of intermediate-ionisation species (S$^{2+}$, Ar$^{2+}$, and Cl$^{2+}$), following the recommendations of \citet{mendezDelgado2023b}. For the total abundances, we adopt the ICFs of \citet{Delgado-Inglada:2014}, which are required for all elements except helium, whose total abundance is computed directly as the sum of its ionic abundances. As a consistency check, we also estimated the integrated N/O ratio using the classical approximation $\mathrm{N/O}=\mathrm{N}^{+}/\mathrm{O}^{+}$. This comparison is motivated by \cite{Delgado-Inglada:2015}, who noted that their earlier nitrogen ICF prescription can introduce correlations between the inferred N/O ratio, helium abundance, and degree of ionisation. For the Helix Nebula, the classical approximation gives  $\log(\mathrm{N/O})=-0.32$, in agreement with the value obtained using the ICF prescription of \cite{Delgado-Inglada:2014}. The agreement indicates that our integrated N/O determination, and consequently the evidence for nitrogen enrichment, is insensitive to the choice between these two prescriptions.

We present the integrated results in Tables~\ref{tab:integral_physical_conditions}, \ref{tab:integral_ionic_abundances}, and \ref{tab:integral_total_abundances}, corresponding to the physical conditions, ionic abundances, and total abundances, respectively. In each case, we compare the integrated measurements with the median values derived from the spatially resolved maps.

\begin{table*}
\centering
\caption{Integrated emission-line fluxes of the Helix Nebula measured from the LVM data. Observed fluxes, $F(\lambda)$, are normalised to $F(\mathrm{H}\beta)=100$, and $I(\lambda)$ correspond to reddening-corrected intensities. The column $\epsilon_I$ indicates the relative uncertainty in the line intensity. Notes flag lines affected by blending, telluric emission, or absorption features, which were carefully considered or excluded in the analysis. The absolute $\mathrm{H}\beta$ flux of the integrated spectrum is $F(\mathrm{H}\beta)=5.35\times10^{-10}\,\mathrm{erg\,s^{-1}\,cm^{-2}}$, and the derived extinction coefficient is $c(\mathrm{H}\beta)=0.05$. The observed Balmer decrement corresponds to $\mathrm{H}\beta/\mathrm{H}\alpha = 0.33$.}
\label{tab:integrated_lines}
\begin{tabular}{lccccc}
\hline
$\lambda_0$ [\AA] & Ion & $F(\lambda)/F(\mathrm{H}\beta)$ & $I(\lambda)/I(\mathrm{H}\beta)$ & $\epsilon_I$ [\%] & Notes \\
\hline
3726.03 & \oii    & 213.3 & 221.1 & 1  & -- \\
3728.82 & \oii    & 298.6 & 309.5 & 1  & -- \\
3868.75 & \neiii  & 101.3 & 104.5 & 1  & -- \\
3889.05 & \hei    & 29.9  & 30.9  & 2  & blend with \hi~3889.05 \\
4026.36 & \hei    & 3.7   & 3.8   & 12 & -- \\
4068.60 & \sii    & 1.5   & 1.5   & 21 & -- \\
4101.73 & \hi     & 28.7  & 29.4  & 2  & -- \\
4340.46 & \hi     & 47.2  & 48.0  & 1  & -- \\
4363.21 & \oiii   & 3.0   & 3.1   & 5  & -- \\
4471.68 & \hei    & 6.9   & 6.9   & 4  & -- \\
4685.71 & \heii   & 13.6  & 13.7  & 2  & -- \\
4740.17 & \ariv   & 0.2   & 0.2   & :  & -- \\
4861.32 & \hi     & 100.0 & 100.0 & 1  & -- \\
4958.91 & \oiii   & 194.1 & 193.5 & 1  & -- \\
5006.84 & \oiii   & 571.8 & 569.0 & 1  & -- \\
5197.90 & \ini    & 4.9   & 4.8   & 10 & affected by telluric emission \\
5200.26 & \ini    & 4.7   & 4.6   & 11 & affected by telluric emission \\
5411.09 & \heii   & 0.8   & 0.8   & 24 & -- \\
5517.71 & \cliii  & 0.4   & 0.4   & 29 & -- \\
5537.88 & \cliii  & 0.2   & 0.2   & :  & -- \\
5754.64 & \nii    & 6.5   & 6.3   & 3  & -- \\
5875.66 & \hei    & 21.1  & 20.5  & 2  & -- \\
6300.30 & \oi     & 182.4 & 175.6 & 1  & affected by telluric emission \\
6312.10 & \siii   & 0.4   & 0.3   & 31 & -- \\
6363.33 & \oi     & 58.7  & 56.5  & 4  & affected by telluric emission \\
6547.48 & \nii    & 178.4 & 171.0 & 1  & -- \\
6562.80 & \hi     & 298.8 & 286.3 & 1  & -- \\
6583.46 & \nii    & 544.7 & 521.8 & 1  & -- \\
6678.15 & \hei    & 6.0   & 5.8   & 2  & -- \\
6716.44 & \sii    & 17.1  & 16.3  & 1  & -- \\
6730.82 & \sii    & 12.3  & 11.8  & 1  & -- \\
7065.71 & \hei    & 3.4   & 3.3   & 6  & -- \\
7135.80 & \ariii  & 26.8  & 25.4  & 1  & -- \\
7281.35 & \hei    & 1.2   & 1.2   & 6  & affected by telluric emission \\
7318.92 & \oii    & 5.2   & 4.9   & 12 & blend with \oii $\lambda~7319.99$ \\
7751.10 & \ariii  & 11.7  & 11.0  & 8  & affected by telluric emission \\
9014.91 & \hi     & 1.5   & 1.4   & :  & affected by telluric absorption/emission \\
9068.06 & \siii   & 8.8   & 8.2   & 5  & -- \\
9228.26 & \hi     & 2.4   & 2.2   & 28 & -- \\
9530.60 & \siii   & 20.5  & 18.9  & 14 & affected by telluric absorption \\
9545.97 & \hi     & 4.4   & 4.1   & 15 & affected by telluric absorption/emission \\
\hline
\end{tabular}
\label{tab:emission_lines}
\end{table*}

\begin{table*}
\centering
\caption{Electron densities and temperatures derived from different diagnostic line ratios. Integrated values correspond to measurements obtained from the total spectrum, while map values correspond to the median and dispersion shown in the spatially resolved maps.}
\begin{tabular}{lcc}
\hline
\multicolumn{3}{c}{\textit{Electron density}} \\
\hline
Diagnostic & Integrated & Median of the map \\
 & ($\mathrm{cm^{-3}}$) & ($\mathrm{cm^{-3}}$) \\
\hline
\oii~$\lambda3726/\lambda3729$ & $70^{+10}_{-20}$ & $47^{+94}_{-32}$ \\
\sii~$\lambda6716/\lambda6731$ & $45^{+10}_{-10}$ & $74^{+177}_{-52}$ \\
\cliii~$\lambda5517/\lambda5538$ & -- & $741^{+2647}_{-579}$ \\
\hline
\multicolumn{3}{c}{\textit{Electron temperature}} \\
\hline
Diagnostic & Integrated & Median of the map \\
 & (K) & (K) \\
\hline
$T_{\rm e}(\text{\oiii}~\lambda4363/\lambda5007)$ & $9400^{+150}_{-150}$ & $8700\pm700$ \\
$T_{\rm e}(\text{\nii}~\lambda5755/\lambda6584)$  & $9200^{+100}_{-100}$ & $9400\pm950$ \\
$T_{\rm e}(\text{\siii}~\lambda6312/\lambda9069)$ & $7700^{+900}_{-900}$ & $9650\pm2214$ \\
$T_{\rm e}(\text{\oii}~\lambda7320/(\lambda3726+\lambda3729))$ & $9650^{+600}_{-700}$ & $10100\pm1600$ \\
$T_{\rm e}(\text{\sii}~\lambda4069/(\lambda6716+\lambda6731))$ & $12000^{+2200}_{-2100}$ & $10050\pm2200$ \\
$T_{\rm e}(\text{\hei}~\lambda5876/\lambda7281)$ & $11800^{+1500}_{-1400}$ (upper limit) & -- \\
\hline
\end{tabular}
\label{tab:integral_physical_conditions}
\end{table*}

\begin{table}
\centering
\caption{Ionic abundances derived from collisionally excited and recombination lines. Values are expressed as $12+\log(X^{i}/\mathrm{H})$. Integrated values correspond to measurements from the total spectrum, while map values correspond to the medians shown in the spatially resolved abundance maps.}
\begin{tabular}{lcc}
\hline
Ion & Integrated & Median of the map \\
 & $12+\log(X^{i}/\mathrm{H})$ & $12+\log(X^{i}/\mathrm{H})$ \\
\hline
He$^{+}$      & $11.17^{+0.01}_{-0.01}$ & $11.14 \pm 0.11$ \\
He$^{2+}$     & $10.04^{+0.01}_{-0.01}$ & $9.81 \pm 0.57$ \\
O$^{+}$       & $8.44^{+0.03}_{-0.02}$ & $8.36 \pm 0.29$ \\
O$^{2+}$      & $8.41^{+0.02}_{-0.02}$ & $8.34 \pm 0.18$ \\
N$^{+}$       & $8.12^{+0.02}_{-0.02}$ & $8.08 \pm 0.29$ \\
S$^{+}$       & $5.94^{+0.02}_{-0.01}$ & $5.87 \pm 0.29$ \\
S$^{2+}$      & $6.38^{+0.18}_{-0.11}$ & $6.05 \pm 0.26$ \\
Cl$^{2+}$     & $5.02^{+0.36}_{-0.18}$ & $4.85 \pm 0.24$ \\
Ar$^{2+}$     & $6.67^{+0.21}_{-0.11}$ & $6.37 \pm 0.20$ \\
Ne$^{2+}$     & $8.26^{+0.03}_{-0.03}$ & $8.17 \pm 0.18$ \\
\hline
\end{tabular}
\label{tab:integral_ionic_abundances}
\end{table}

\begin{table}
\centering
\caption{Total elemental abundances derived from the integrated spectrum and from the spatially resolved maps. Values are given as $12+\log(X/\mathrm{H})$. Map values correspond to the medians shown in the abundance maps.}
\begin{tabular}{lcc}
\hline
Element & Integrated & Median of the map \\
 & $12+\log(X/\mathrm{H})$ & $12+\log(X/\mathrm{H})$ \\
\hline
He & $11.20^{+0.01}_{-0.01}$ & $11.12 \pm 0.04$ \\
O  & $8.74^{+0.02}_{-0.02}$ & $8.74 \pm 0.14$ \\
N  & $8.42^{+0.03}_{-0.03}$ &  --\\
S  & $6.53^{+0.11}_{-0.11}$ & $6.31 \pm 0.19$ \\
Cl & $5.15^{+0.28}_{-0.27}$ & -- \\
Ar & $6.71^{+0.15}_{-0.15}$ &  --\\
Ne & $8.28^{+0.03}_{-0.03}$ &  --\\
\hline
\end{tabular}
\label{tab:integral_total_abundances}
\end{table}

The integrated spectrum provides a complementary perspective to the spatially resolved analysis, but its interpretation requires some care. In principle, the use of integrated line ratios in a strongly stratified nebula such as the Helix may introduce biases, since the observed fluxes correspond to emissivity-weighted quantities rather than simple averages of the underlying physical conditions, and therefore may emphasise the brightest or hottest emitting zones. This effect is particularly relevant for temperature-sensitive CELs, whose emissivities depend exponentially on $T_{\rm e}$.

In this context, the comparison between the integrated results and the spatially resolved maps is particularly instructive. The oxygen abundance derived from the integrated spectrum, $12+\log(\mathrm{O/H})=8.74$ (Table~\ref{tab:integral_total_abundances}), is remarkably close to the median value obtained from the spatially resolved analysis, despite the fact that the temperature maps exhibit apparent variations of several thousand kelvin across the plane of the sky (see Figs.~\ref{fig:Te_OIII}--\ref{fig:Te_SII}).

If the observed spatial variations in $T_{\rm e}$ from Fig.~\ref{fig:Radial_Te} reflect genuine thermal inhomogeneities, then the integrated temperature represents a luminosity-weighted value that overemphasises hotter regions. In this case, the derived abundances from CELs would be systematically biased low, as predicted in the classical framework of temperature fluctuations \citep{Peimbert1967}. On the other hand, if part of the apparent temperature structure is dominated by recombination contributions to auroral lines, then the inferred temperatures are not purely thermal diagnostics. In such a scenario, the resulting bias does not disappear in the integrated spectrum: the same excess auroral emission is included in the total flux, and therefore the integrated $T_{\rm e}$ and abundances would be affected in a similar way as in the spatially resolved case. In other words, the integrated spectrum averages the bias rather than eliminating it.

These two scenarios differ in their physical origin, but can lead to qualitatively similar systematic effects on CEL-based abundances in integrated spectra. In the case of genuine temperature fluctuations, the bias arises from the exponential dependence of CEL emissivities on $T_{\rm e}$, whereas recombination contributions scale approximately linearly with the abundance of the recombining ion. The observed consistency between integrated abundances and the median values from the maps may therefore suggest that at least part of the apparent temperature structure does not necessarily reflect true spatial variations in the electron temperature. Instead, it could be partly produced by recombination contributions to the auroral lines used to derive $T_{\rm e}$. It is important to note that this effect is not limited to the auroral transitions themselves —where it is proportionally larger due to their intrinsic weakness— but may also affect other lines, particularly \oii~$\lambda\lambda3726,3729$, which play a key role in abundance determinations \citep{Stasinska:2005}. 
The good agreement between the integrated abundances and the spatially resolved medians should therefore not be interpreted as evidence for the absence of systematic effects, nor as an indication that spatially resolved spectroscopy is unnecessary. On the contrary, it suggests that the intrinsic biases affecting CEL diagnostics may be present in a comparable way across the nebula, and that only wide-field, spatially resolved observations can provide the necessary leverage to disentangle their origin.

\section{Discussion}
\label{sec:discussion}
\subsection{Temperature variations}
\label{sec:temp_variations}
Recent spatially resolved studies have shown that the connection between CEL--RL abundance discrepancies and temperature variations may differ from object to object. For example, \citet{Singh:2026} reported a strong correlation between the abundance discrepancy and temperature variations in M8. On the other hand, \citet{Sattler:2026} found no significant temperature biases in M20, despite the detection of \oiirls~RLs and the presence of abundance discrepancies in previous works \citep{Garcia-Rojas:2006}. These contrasting results suggest that differences in metallicity, ionisation structure, radiation hardness, density, and geometry among ionised nebulae may play a key role in shaping the observed temperature variations and their connection to the abundance discrepancy problem. In this context, the spatially resolved temperature structure of the Helix provides an opportunity to investigate which physical mechanisms may produce such variations within an evolved PN.

A particularly noteworthy result of the spatially complete temperature mapping is the contrasting radial behaviour of the different diagnostics (Fig.~\ref{fig:Radial_Te}). The high-ionisation $T_{\rm e}(\text{\oiii})$ profile exhibits a characteristic U-shaped morphology, decreasing from approximately 10,000\,K in the central region to $\sim8400\,K$ at intermediate radii before increasing to $\sim9850\,K$ near $r \sim 0.42$\,pc. In contrast, $T_{\rm e}(\text{\nii})$ increases monotonically toward the outer nebula, whereas $T_{\rm e}(\text{\oii})$ decreases outward. A qualitatively similar high--low--high pattern has been reported in NGC~6543 \citep{Hyung2001, Wesson2004}. The opposite behaviour of the low- and high-ionisation diagnostics also shows that they cannot be interpreted simply as measurements of a single radial temperature field, since their emissivities preferentially weight different ionisation zones along the line of sight.

An important constraint on the origin of the central $T_{\rm e}(\text{\oiii})$ enhancement is provided by the present-day stellar wind. No fast wind is currently detected from the central star of the Helix \citep{Meaburn2005}, making ongoing stellar-wind energy injection an unlikely explanation for the elevated central temperature. We therefore favour an interpretation in which the central-to-intermediate variation of $T_{\rm e}(\text{\oiii})$ primarily reflects the thermal and ionisation stratification of the nebula, modulated by its three-dimensional geometry and line-of-sight projection, rather than requiring current mechanical energy input from the central star.  The outer rise in $T_{\rm e}(\text{\oiii})$ may likewise reflect changes in the ionisation and thermal balance toward the nebular boundary, although the present observations do not allow these effects to be separated uniquely from projection.

The monotonically increasing $T_{\rm e}(\text{\nii})$ profile is more difficult to interpret. Its radial behaviour is opposite to the outward decrease in $T_{\rm e}(\text{\oii})$, despite both diagnostics preferentially tracing lower-ionisation gas. This difference reinforces the conclusion that the various temperature diagnostics sample distinct emitting volumes and respond differently to ionisation stratification and projection effects. At the same time, the persistent outward increase in the thermal structure of \nii may indicate that photoionisation alone cannot describe the behaviour of the low-ionisation gas. The recently identified bow shocks around the Helix provide independent evidence that interaction between earlier nebular ejecta and the ambient ISM is occurring at larger radii \citep{VanDokkum2026}, although these structures do not directly demonstrate that such interaction is responsible for the observed $T_{\rm e}(\text{\nii})$ gradient. Additional low-level mechanical heating associated with the shell expansion or interaction with the surrounding material therefore remains a plausible possibility. A similar interpretation has been proposed for NGC 6543, where photoionisation and spectral hardening alone were unable to reproduce the elevated temperature measured in its outer regions and additional shock heating was invoked \citep{Hyung2001}.

The spatially resolved velocity field provides useful context for this interpretation, showing that the outer low-ionisation material forms part of a dynamically complex expanding structure. However, the observed kinematics do not by themselves demonstrate that shocks are responsible for the temperature increase. The Helix has a toroidal and multi-component geometry, and line-of-sight projection through this structure may also contribute to the radial temperature trends. The present observations therefore do not allow a unique separation between photoionisation, projection effects, and possible low-level mechanical heating.

\subsection{The apparent oxygen deficit and unobserved O$^{3+}$}
\label{sec:oxygen_deficit}
\begin{figure}
    \centering
    \includegraphics[width=1.0\linewidth]{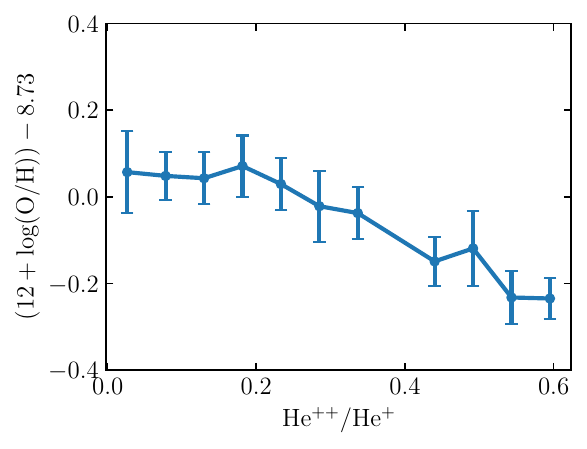}
\caption{Apparent oxygen deficit relative to the nebular median abundance as a function of the He$^{2+}$/H$^{+}$ abundance computed in spatial bins across the nebula. A clear positive correlation is observed, with regions of stronger He$^{2+}$ emission showing systematically lower apparent O/H values. Since He$^{2+}$ and O$^{3+}$ require similarly hard ionising photons, this behaviour supports the interpretation that the central apparent oxygen deficit is caused by oxygen hidden in higher ionisation stages not directly observed in the optical.}
\label{fig:OvsHe}
\end{figure}

In Section~\ref{sec:chemical_abundances}, we identified an apparent oxygen deficit toward the central region of the Helix Nebula. We briefly mentioned that this could be due to a shift from O$^{2+}$ to O$^{3+}$, which is not directly observable in our optical data. Additional support for this interpretation is provided by the spatial distribution of He$^{2+}$, traced by the strong \heii~$\lambda4686$ recombination line. The production of He$^{2+}$ requires photons above 54.4 eV, very similar to those needed to produce O$^{3+}$. Therefore, regions exhibiting strong \heii~emission are natural candidates to host significant O$^{3+}$ fractions. Figure~\ref{fig:OvsHe} compares the apparent oxygen deficit relative to the nebular median with the He$^{2+}$/H$^{+}$ abundance, revealing a clear correlation: spaxels with stronger He$^{2+}$ emission systematically show lower apparent O/H values. This behaviour strongly suggests that the missing oxygen in the central regions is not physically absent, but rather hidden in higher ionisation stages not directly measured in the optical, as the strongest transitions of O$^{3+}$ lie in the ultraviolet and infrared wavelength ranges.

Indeed, the presence of diffuse \oiv{} \(25.9\,\mu\)m 
line emission from the centre of the Helix was already suspected from IRAS imaging \citep{Leene1987}
and this was confirmed by Spitzer spectroscopy \citep{Su2007}.
The Helix is thus an example of Group~3 in the mid-infrared PN classification of \citet{Chu2009}:
a population of highly evolved nebulae that show centrally condensed \(24\,\mu\)m emission \citep{Garcia-Diaz2018},
due to \oiv{} and (in the case of hotter central stars) \nev{} lines \citep{Flagey2011}.

\begin{figure}
    \centering
    \includegraphics[width=1.0\linewidth]{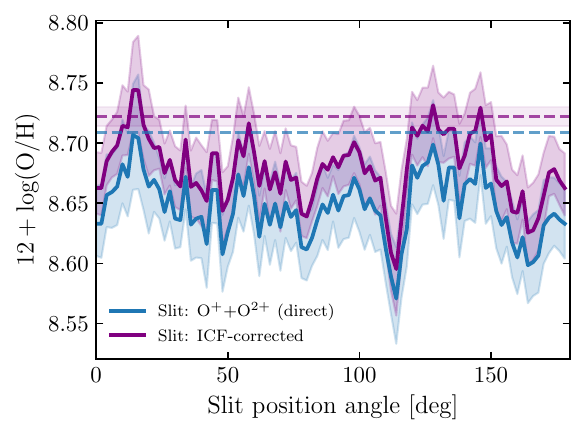}
    \caption{O/H abundance as a function of the position angle of simulated long-slit apertures extracted from the Helix Nebula LVM data cube. The blue curve shows the O/H abundance derived from the directly observed ionic abundances, O$^{+}$ + O$^{2+}$, whereas the purple curve shows the abundance after applying the oxygen ICF of \citet{Delgado-Inglada:2014}. The corresponding dashed horizontal lines indicate the O/H values obtained using the full spatial coverage of the LVM data set. The shaded regions indicate the 16th--84th percentile uncertainty intervals.}
    \label{fig:OH_slits}
\end{figure}

The spatially complete LVM observations allow us to quantify an additional effect that cannot be addressed straightforwardly with traditional long-slit observations. Namely, the sensitivity of the oxygen abundance to spatial sampling. To investigate this effect, we simulated long-slit apertures across our Helix observations at different position angles and repeated the oxygen-abundance analysis for each aperture using the same ionic diagnostic and the \cite{Delgado-Inglada:2014} ICF prescription adopted for the full IFU data set. Figure~\ref{fig:OH_slits} shows that the inferred O/H abundance varies appreciably with the slit position. Although some slit orientations recover values close to those obtained from the complete IFU coverage, the slit-based abundances are generally lower, even after the ICF correction. Thus, the use of an appropriate ICF does not necessarily eliminate the bias introduced by incomplete spatial sampling. In a strongly ionisation-stratified nebula, the ionic fractions entering the correction are themselves aperture dependent because different slit positions sample different proportions of the O$^{+}$ and O$^{2+}$, and higher-ionisation zones. The principal advantage of the spatially complete IFU coverage is therefore not the identification of the ionisation stratification itself, which is well established for the Helix, but the ability to quantify this effect on the abundance determination and to obtain a more representative nebula-integrated O/H abundance.

This aperture dependence also helps explain part of the scatter observed in the total oxygen map. Since the relative importance of O$^{3+}$ is expected to vary smoothly with local ionisation parameter, geometry, and line-of-sight projection effects, any abundance estimate based solely on O$^{+}$ and O$^{2+}$ can introduce artificial spatial structure in highly ionised regions. The three-dimensional geometry of the Helix Nebula is particularly relevant in this regard. Several studies have described the object as a bipolar or thick toroidal structure viewed nearly pole-on \citep{Meaburn1998,ODell1998,ODell2004,Meaburn2005}. Although the overlap of these zones may be less severe than in a spherical shell, projection effects still influence how the ionisation structure is mapped onto the plane of the sky and, consequently, the ionic fractions from which the total abundance is inferred.

\subsection{The oxygen abundance of the Helix in context}
Overall, the oxygen abundance structure of the Helix Nebula is consistent with a chemically homogeneous nebula of apparently Solar oxygen composition, whose central abundance deficit is primarily driven by ionisation stratification and the presence of unobserved higher ionisation stages. However, it is important to note that this apparently Solar oxygen abundance, while fully consistent with expectations for an object located in the Solar neighbourhood, is also intriguingly anomalous when compared with many nebular abundance studies based on the so-called ``direct method'', the same general approach adopted here. Analyses of photoionised nebulae in the local Galactic environment, including both \hii~regions and planetary nebulae \citep{Perinotto:2006,MendezDelgado:2022}, often yield oxygen abundances lower than the Solar value by approximately $0.2$ dex when temperature fluctuations are not considered. One of the clearest examples is the Orion Nebula, the nearest massive star-forming region to the Sun ($\sim420$ pc), for which direct-method abundances around $12+\log(\mathrm{O/H})\approx8.50$ are commonly reported \citep{Esteban:2004}. A better overall consistency between stellar and nebular abundances is generally recovered when either heavy-element recombination lines are used or when a correction for temperature inhomogeneities is applied through the mean-square temperature fluctuation parameter $t^{2}$ introduced by \citet{Peimbert1967} (\citealt{MartinezHernandez:2026}, Lugo et al., in prep.; López-Valdivia et al., in prep.).

Indeed, applying a correction for temperature fluctuations within the line of sight, \citet{Peimbert:1987} derived an oxygen abundance of $12+\log(\mathrm{O/H})=8.89$ for the Helix, approximately $0.2$ dex higher than the value inferred in this work. If the relative oxygen enhancement of the Helix with respect to other nearby nebulae were instead due to stellar migration, the progenitor star would need to have formed significantly closer to the Galactic centre. Assuming a typical Galactic oxygen abundance gradient \citep{MendezDelgado:2022}, an offset of $\sim0.2$ dex would correspond to a radial displacement of roughly $4$--$5$ kpc. This interpretation is valid in the context of radial-migration models. Simulations of galaxy disks have shown that a substantial fraction of stars currently located in a solar-neighbourhood-like annulus may have formed at different Galactocentric radii. In particular, \citet{Rovskar2008} found that metal-rich stars in the solar neighbourhood can originate over a broad range of birth radii. Therefore, if the relative high oxygen abundance of the Helix reflects the initial compositions of its progenitor, rather than abundance-diagnostic effects, it could in principle indicate formation at a smaller Galactocentric radius followed by outward migration.

An alternative possibility is genuine oxygen enrichment by the progenitor star itself, as proposed by \citet{Delgado-Inglada:2015} for some planetary nebulae, particularly those associated with carbon-rich dust chemistry. Whether the explanation involves migration, stellar nucleosynthesis, temperature fluctuations, or a combination of these effects, the Helix Nebula illustrates the important role that planetary nebulae may play in the chemical feedback from low- and intermediate-mass stars into the interstellar medium.

Surveys of planetary nebulae in nearby galaxies have already demonstrated the value of PNe as tracers of chemical enrichment, stellar nucleosynthesis, and galaxy evolution \citep[e.g.][]{Hernandez-Martinez2011,Garcia-Rojas2016, Tsakonas2025}. However, in such extragalactic samples, individual nebulae are generally unresolved, making it difficult to connect their integrated abundances with the underlying spatial variations in ionisation structure, temperature, and chemical composition. The SDSS-V LVM opens a complementary observational window in this context, enabling sub-parsec studies within the Milky Way together with wide-field spatial coverage. This allows nearby objects such as the Helix Nebula to be used as spatially resolved benchmarks for understanding how evolved low- and intermediate-mass stars contribute chemically processed material to the interstellar medium.

\subsection{Progenitor enrichment and abundance anomalies}
The abundance-ratio maps provide insight into both the chemical imprint of the progenitor star and the limitations of interpreting ionic ratios as direct elemental abundance tracers. The observed helium and nitrogen enrichment indicates that the progenitor experienced chemical processing during its post-main-sequence evolution. In particular, the enhanced He/H ratio is consistent with the general behaviour of observed Galactic PNe and with stellar evolution models in which the dredge-up episodes increase the surface helium abundance \citep{Henry2018}.
The nitrogen enrichment observed in the Helix is also consistent with the general trend found among Galactic PNe. \cite{Henry2018} showed that enhanced N/O ratios are observed over a relatively broad range of inferred initial masses, including objects with initial masses of 2 -- 4\,M$_{\odot}$. However, the interpretation of N/O in terms of progenitor mass is strongly model dependent. In particular, the available nucleosynthesis models do not fully reproduce the observed N/O distribution, especially at low progenitor masses, which may indicate the contribution of additional mixing processes such as rotation-induced mixing \citep{Henry2018}.
\cite{Ventura2018} further demonstrated that the predicted surface abundance pattern depends sensitively on the treatment of convection and mixing during the AGB phase. In their solar-metallicity ATON models, HBB becomes efficient at initial masses of approximately 3.5\,M$_{\odot}$, producing a sharp increase in N/O together with strong carbon depletion, with final N/O $\sim$\, 1.5 -- 2.0 and C/O $<$ 0.1. By contrast, the MONASH models predict the onset of HBB at initial masses approximately 1\,M$_{\odot}$ higher and retain substantially larger C/O ratios at intermediate and high masses. These differences arise primarily from the adopted treatment of convection and the resulting efficiency of HBB and third dredge-up.
In this context, the abundance pattern of the Helix, with moderate enhancements of He and N and the previously reported C/O $\sim$\,0.9 \citep{Henry-etal-1999}, is broadly consistent with nucleosynthetic processing in a low to intermediate-mass AGB progenitor. However, these abundances do not uniquely constrain the initial stellar mass or demonstrate the occurrence of HBB, because the predicted C/O and N/O ratios, as well as the mass threshold for HBB, depend strongly on the adopted stellar-evolution prescription.

The behaviour of sulfur is particularly noteworthy. The ICF-corrected S/Ar ratio is systematically below the Solar reference by approximately one dex over most radii. Since sulfur and argon are both $\alpha$-elements and are not expected to be significantly modified during the evolution of low- and intermediate-mass stars \citep{Esteban:2025}, such a low value cannot be straightforwardly explained by stellar nucleosynthesis. Instead, it is consistent with the well-known sulfur anomaly in planetary nebulae \citep{Henry:2004,Shingles:2013,Tan:2024}, namely the tendency for sulfur abundances derived from nebular spectroscopy to appear lower than expected when compared with other $\alpha$-elements, particularly oxygen and argon. The S/Ar ratio derived from the integrated spectrum, shown as a coloured dotted line in Fig.~\ref{fig:radial_relative_abundances}, is consistent with the low values traced by the spatially resolved profile, supporting the interpretation that the sulfur deficit is a global property of the nebula.

The fact that the S/Ar deficit persists after applying the model-guided ICFs suggests that it is unlikely to be caused solely by the particular ionic stages observed in the optical. Nevertheless, given the approximate nature of the adopted corrections, the absolute level of the sulfur deficit should be interpreted with some caution. Additional support for a sulfur deficit is provided by the S/Cl ratio. Although the \cliii~lines are detected in only a limited number of spaxels, Cl$^{2+}$ provides a useful reference because its ionisation structure is expected to broadly resemble that of S$^{2+}$. The relevant ionisation potential ranges are similar: S$^{+}\rightarrow$S$^{2+}$ requires 23.3 eV and S$^{2+}\rightarrow$S$^{3+}$ 34.8 eV, while Cl$^{+}\rightarrow$Cl$^{2+}$ requires 23.8 eV and Cl$^{2+}\rightarrow$Cl$^{3+}$ 39.6 eV. Therefore, both ions should coexist over much of the same nebular volume. The sub-Solar S/Cl values, despite their larger uncertainties, are therefore qualitatively consistent with the S/Ar result and support the interpretation that the sulfur deficit is not purely an ionisation artefact. The integrated S/Cl ratio shows the same sub-Solar behaviour, providing an additional global consistency check.

The inclusion of Cl/O and Ar/O in Figure~\ref{fig:radial_relative_abundances} provides important context for this difference. Ne/O remains comparatively close to the Solar reference over most of the nebula, whereas Ar/O shows a systematic decrease toward larger radii and Cl/O is sub-Solar over the radial range where it can be reliably measured, albeit with comparatively large uncertainties. Thus, Cl and Ar do not behave as completely interchangeable reference elements in the spatially resolved Helix data. In particular, the offset of approximately 0.5\,dex between S/Cl and S/Ar is comparable to the corresponding difference between Cl/O and Ar/O. This follows directly from the abundance ratios, since the difference between S/Ar and S/Cl is determined by the relative Cl/Ar abundance. The ratios inferred from the integrated spectrum show a qualitatively similar relative behaviour, reinforcing the interpretation that the different sulfur deficits obtained with Cl and Ar partly reflect the behaviour of the reference elements themselves. The different amplitudes of the sulfur deficit inferred using Cl and Ar therefore do not necessarily imply inconsistent sulfur measurements, but partly reflect the different behaviour of the reference elements themselves.

This comparison also provides a useful test of whether Cl and Ar can be regarded as more robust metallicity tracers than oxygen in the Helix. Our spatially resolved results do not support such a simple interpretation. While Ne/O remains relatively stable, both Cl/O and Ar/O show spatial variations, indicating that the inferred ratios remain sensitive to ionisation structure, spatial sampling, and the corresponding abundance corrections. Nevertheless, sulfur remains systematically deficient relative to both Ar and Cl, so the presence of the sulfur anomaly itself is robust, even though its inferred magnitude depends on the reference element adopted.

An additional explanation proposed in the literature is that part of the sulfur may be depleted onto dust grains or locked into sulfur-bearing solid compounds such as MgS, FeS, or other refractory condensates \citep[e.g.][]{Henry:2012,Delgado-Inglada:2014}. If this were the dominant mechanism in the Helix Nebula, however, the associated dust component would need to be at least partially decoupled from the material traced by the optical extinction. As shown in our Fig.~\ref{fig:extinction_map}, the Helix exhibits very low internal attenuation over most of the nebula, with only modest enhancements in the outer ring. This implies that any sulfur-bearing dust reservoir large enough to explain a depletion of nearly one dex would either have to possess a low extinction efficiency at optical wavelengths, be concentrated in dense clumps with small covering fraction, or reside in phases not effectively sampled by the line-of-sight extinction diagnostics. In this sense, while dust depletion may contribute to the sulfur anomaly, the weak overall extinction observed in the Helix suggests that it is unlikely to provide a complete explanation on its own.

\subsection{Integrated versus spatially resolved abundance determinations}
The integrated spectrum extracted over the full LVM footprint provides an important complementary perspective. Because it incorporates the complete ionisation structure of the nebula, it is especially suitable for the application of classical ionisation correction factor prescriptions calibrated from photoionisation models. In this sense, the integrated Helix spectrum constitutes a valuable local benchmark for assessing how well aperture-integrated spectra recover nebular physical conditions and abundances when compared with spatially resolved measurements. This is particularly relevant for unresolved extragalactic planetary nebulae, where spatially integrated observations are generally the only available information.

The integrated spectrum of the Helix Nebula provides a complementary view to the spatially resolved analysis, but its interpretation must account for the emissivity weighting inherent to CEL diagnostics. In a strongly stratified nebula, integrated line ratios preferentially weight regions according to their local physical conditions, and therefore spatially averaged and integrated quantities are not expected to be strictly equivalent. This is particularly relevant for temperature-sensitive auroral lines, because regions with higher inferred electron temperatures can contribute disproportionately to the integrated CEL ratios. If the spatial variations in $T_{\rm e}$ reflected genuine thermal inhomogeneities, the integrated spectrum could therefore be biased toward higher temperatures and correspondingly lower CEL abundances. Recombination provides an alternative complication, since it can enhance some auroral lines and increase the inferred temperatures without requiring equally large true thermal variations. Because this contribution depends on the abundance of the recombining ion, it may affect both spatially resolved and integrated measurements. The relatively good agreement between the oxygen abundance derived from the integrated spectrum and the median value obtained from the spatially resolved maps is therefore noteworthy. However, this agreement alone cannot be used as evidence for a significant recombination contribution to the auroral lines.

Quantifying the recombination contributions observationally is difficult. As discussed by \cite{Wesson2018}, recombination can affect the \oii~density diagnostic while leaving \sii~comparatively unaffected, so a systematic discrepancy between densities derived from these two diagnostics can provide an indication of such an effect. In our data, however, the integrated and spatially representative density estimates do not provide a clear and consistent signature of this behaviour. Consequently, recombination should be regarded as a possible secondary effect rather than as an explanation for the agreement between the integrated and spatially resolved abundances, and its impact on the derived abundances is likely to be small.

\section{Summary and Conclusions}
\label{sec:conclusions}
We have presented the first spatially resolved integral-field spectroscopic analysis of the physical and chemical structure of the Helix Nebula (NGC~7293) based on observations obtained with the SDSS-V Local Volume Mapper (LVM). Owing to its wide field of view, contiguous spatial sampling, and broad wavelength coverage, these data provide a global spectroscopic view of one of the nearest planetary nebulae, overcoming the strong aperture limitations inherent to previous long-slit observations. By reconstructing datacubes and analysing the nebular emission on a spaxel-by-spaxel basis, we derived maps of extinction, electron density, electron temperature, ionisation structure, and chemical abundances across the nebula.

The extinction structure confirms that the Helix is a remarkably transparent object. The central cavity exhibits very low reddening, while moderate increases in $c(\mathrm{H}\beta)$ are observed toward the bright outer ring. This morphology is consistent with the projected toroidal shell inferred in previous optical and infrared studies, where dust is concentrated around the evacuated inner cavity rather than uniformly mixed throughout the nebula.

The density structure indicates that the Helix is predominantly a low-density nebula. Diagnostics based on \sii~, \oii~, and \cliii~ yield lower values than $n_{\rm e}\sim10^{2}\ \mathrm{cm}^{-3}$ over most of the mapped area, with only mild enhancements toward the inner shell. Much of the apparent small-scale scatter is attributable to the limited sensitivity of classical optical diagnostics in the low-density regime, where observational uncertainties become comparable to the density dependence of the line ratios themselves.
By contrast, the temperature structure is highly non-uniform. The high-ionisation indicator $T_{\rm e}(\text{\oiii}~\lambda4363/\lambda5007)$ exhibits a characteristic U-shaped radial behaviour, with elevated temperatures toward both the central cavity and the outer nebular rim, and a minimum at intermediate radii. The low-ionisation tracers $T_{\rm e}(\text{\nii}~\lambda5755/\lambda6584)$ and $T_{\rm e}(\text{\oii}~\lambda7320/(\lambda3726+\lambda3729))$ show distinct but complementary behaviours, revealing that the thermal structure depends strongly on the ionisation zone being probed. The central enhancement in $T_{\rm e}(\text{\oiii})$ is most naturally associated with the thermal and ionisation stratification produced by the hard radiation field of the central star, modulated by the three-dimensional geometry and line-of-sight projection of the nebula. The outward increase observed in some temperature diagnostics may likewise reflect changes in the ionisation and thermal balance toward the nebular boundary, although additional low-level mechanical heating associated with shell expansion or interaction with the surrounding material cannot be excluded.

These results provide new spatially resolved insight into the long-standing temperature fluctuations problem. Since \citet{Peimbert1967}, temperature inhomogeneities have been considered as a possible contributor to the systematic differences between abundances derived from CELs and those obtained from heavy-element RLs. Although the present data do not provide a direct measurement of this abundance discrepancy, the markedly different radial behaviours of the temperature diagnostics demonstrate that a single characteristic temperature cannot fully describe the thermal structure of the Helix. At the same time, we note that some fraction of the apparent behaviour of specific auroral diagnostics may also be influenced by recombination contributions or related processes, highlighting the need for caution when interpreting single-temperature indicators.

The ionic abundance maps recover the strong ionisation stratification already known in the Helix. High-ionisation species such as He$^{2+}$ dominate the inner regions, while lower ionisation stages become progressively more important toward the bright outer shell. The main advantage of the spatially contiguous LVM coverage is therefore not the identification of this stratification itself, but the possibility of quantifying how it affects the physical conditions and chemical abundances inferred across the nebula.

The oxygen abundance derived from the directly observed O$^{+}$ and O$^{2+}$ ions indicates a chemically homogeneous nebula of approximately Solar composition, with a representative median value of $12+\log(\mathrm{O/H})\approx8.73$. Lower apparent O/H values in the innermost regions correlate strongly with enhanced He$^{2+}$/H$^{+}$ emission, implying that a non-negligible fraction of oxygen is shifted into higher ionisation stages, primarily O$^{3+}$, not directly measured in the optical. The central ``oxygen deficit'' is therefore best interpreted as an ionisation effect rather than a true abundance variation. More importantly, our simulated long-slit experiment demonstrates that the inferred O/H abundance depends on the position and orientation of the spectroscopic aperture and can remain biased even after applying an ionisation correction. Thus, the spatially complete IFU coverage provides a more representative determination of the nebular oxygen abundance and directly quantifies the aperture effects that can affect traditional long-slit measurements of strongly stratified planetary nebulae.

The derived helium abundance, $12+\log(\mathrm{He/H})\approx11.12$, together with an enhanced N/O ratio, is consistent with chemical processing in the progenitor star during its post-main-sequence evolution. Together with the previously reported C/O $\sim0.9$, this abundance pattern is broadly consistent with nucleosynthetic processing in a low- to intermediate-mass AGB progenitor. However, these abundances do not uniquely constrain the initial stellar mass or demonstrate the occurrence of HBB, since the predicted C/O and N/O ratios and the mass threshold for HBB depend strongly on the adopted stellar-evolution prescription.

Our sulfur results are also noteworthy. Relative abundance ratios involving sulfur remain systematically low when compared with oxygen, argon, and chlorine, suggesting that the Helix may display the well-known sulfur anomaly reported for many planetary nebulae. This behaviour is unlikely to be explained solely by ionisation effects, and may instead reflect a combination of missing higher ionisation stages, dust depletion, and/or unresolved issues in sulfur atomic data.

Overall, our results demonstrate that spatial completeness is essential for deriving reliable physical conditions and abundances in extended ionised nebulae. Measurements based on selected apertures may be significantly biased by ionisation stratification, temperature gradients, and projection effects. Wide-field IFU observations such as those provided by LVM overcome these limitations and open a new avenue for precision studies of spatially extended nebulae, including PNe and Galactic H II regions.

The Helix Nebula therefore emerges as a key laboratory for testing nebular diagnostics, ionisation correction factors, and the physical origin of abundance discrepancies in photoionised plasmas. Future deeper LVM observations, combined with ultraviolet and infrared spectroscopy, will be particularly valuable for directly constraining ions such as O$^{3+}$ and S$^{3+}$, as well as for detecting heavy-element recombination lines and further clarifying the thermal and chemical structure of this iconic nebula.

\section*{Data availability}
The LVM survey data will be made publicly available with SDSS Data Release 20. MWISP data can be accessed upon formal request through the MWISP portal.

\section*{Acknowledgements} 

The authors thank the anonymous referee for their constructive comments and suggestions, which helped improve the quality of this paper.
ROD and JAT thank support from UNAM DGAPA PAPIIT project IN102324.\\
JEMD, AZLA, LCCC, CM, OA and SFS acknowledge support from the Secretaría de Ciencia, Humanidades, Tecnología e Innovación (SECIHTI) project CBF-2025-I-2048, ``Resolviendo la Física Interna de las Galaxias: De las Escalas Locales a la Estructura Global con el SDSS-V Local Volume Mapper''.\\
JEMD, AZLA, LCCC, CM and SFS acknowledge support from UNAM/DGAPA/PAPIIT/IA103326 project ``DESIRED (DEep Spectra of ionised Regions Database): de las emisiones más sutiles a la física fundamental del universo’’ .\\
LS acknowledges support from PAPIIT UNAM grant IN107625\\
C.R.-Z. acknowledges support from PAPIIT UNAM grant IN107226\\
E.J.J. acknowledges support from the FONDECYT Regular grant number 1262304 and the ANID CATA-BASAL project FB210003.\\
R. Z. would like to express his appreciation for the support provided by SECIHTI (Secretaría de Ciencia, Humanidades, Tecnología e Innovación) in the form of a Postdoctoral Grant.\\
AZLA gratefully acknowledges the support provided by the Postdoctoral Program (POSDOC) of UNAM (Universidad Nacional Autónoma de México)\\
OE acknowledges funding from the Deutsche Forschungsgemeinschaft (DFG, German Research Foundation) -- project-ID 541068876.\\
G.A.B. acknowledges the support from the ANID Basal project FB210003. 
A.S. acknowledges the support from the ANID Basal projects FB21250718 and FB210003.\\
Funding for the Sloan Digital Sky Survey V has been provided by the Alfred P. Sloan Foundation, the Heising-Simons Foundation, the National Science Foundation, and the Participating Institutions. SDSS acknowledges support and resources from the Center for High-Performance Computing at the University of Utah. SDSS telescopes are located at Apache Point Observatory, funded by the Astrophysical Research Consortium and operated by New Mexico State University, and at Las Campanas Observatory, operated by the Carnegie Institution for Science. The SDSS web site is www.sdss.org. 

SDSS is managed by the Astrophysical Research Consortium for the Participating Institutions of the SDSS Collaboration, including Caltech, The Carnegie Institution for Science, Chilean National Time Allocation Committee (CNTAC) ratified researchers, The Flatiron Institute, the Gotham Participation Group, Harvard University, Heidelberg University, The Johns Hopkins University, L’Ecole polytechnique fédérale de Lausanne (EPFL), Leibniz-Institut für Astrophysik Potsdam (AIP), Max-Planck-Institut für Astronomie (MPIA Heidelberg), The Flatiron Institute, Max-Planck-Institut für Extraterrestrische Physik (MPE), Nanjing University, National Astronomical Observatories of China (NAOC), New Mexico State University, The Ohio State University, Pennsylvania State University, Smithsonian Astrophysical Observatory, Space Telescope Science Institute (STScI), the Stellar Astrophysics Participation Group, Universidad Nacional Autónoma de México, University of Arizona, University of Colorado Boulder, University of Illinois at Urbana-Champaign, University of Toronto, University of Utah, University of Virginia, Yale University, and Yunnan University.\\
This work has made an extensive use of NASA's Astrophysics Data System (ADS).

%%%%%%%%%%%%%%%%%%%%%%%%%%%%%%%%%%%%%%%%%%%%%%%%%%

%%%%%%%%%%%%%%%%%%%% REFERENCES %%%%%%%%%%%%%%%%%%

% The best way to enter references is to use BibTeX:

%\bibliographystyle{mnras}
%\bibliography{example} % if your bibtex file is called example.bib

\bibliographystyle{mn2e}
\bibliography{bibliography}

\appendix

% Don't change these lines
\bsp	% typesetting comment
\label{lastpage}
\end{document}